\documentclass{aa}
\usepackage{lineno}
\usepackage[utf8]{inputenc}
\usepackage{natbib}
\usepackage{amsmath}    
\usepackage{subfigure}
\usepackage{graphicx}
\usepackage{float}
\usepackage{natbib}
\usepackage{dcolumn}
\usepackage{bm}
\usepackage{ulem}
\usepackage{color}
\usepackage{txfonts}
\usepackage{rotating}
\usepackage{booktabs}
\usepackage{array}
\usepackage{tabularx}
\usepackage{threeparttable}
\usepackage{diagbox}
\usepackage{multirow}
\usepackage{makecell}
\usepackage{longtable}
\usepackage{mhchem}
\usepackage{afterpage}
\usepackage{xcolor}
\usepackage{setspace}
\usepackage{geometry}
\usepackage[colorlinks=true, allcolors=blue]{hyperref} 
\definecolor{myblue}{RGB}{0,0,255}  
\definecolor{myblack}{RGB}{0,0,0}   
                                
\begin{document}
\nolinenumbers 

\title{The influence of rotation and metallicity on the explodability of massive stars}

\author{Renyu Luo\inst{1}
\and    Chunhua Zhu\inst{1}
\and    Guoliang L\"{u}\inst{1*}
\and    Helei Liu\inst{1}
\and    Sufen Guo\inst{1*}
\and    Lei Li\inst{1}
\and    Zhuowen Li\inst{1}
}
 \institute{School of Physical Science and Technology, Xinjiang University, Urumqi, 830046, China\\
              \email{guolianglv@xao.ac.cn;947559540@qq.com}}

  \abstract
   {During the late stages of massive stellar evolution, failed supernovae (FSN) may form through core-collapse processes. The traditional evaluation criterion $\xi_{2.5}$ $=$ 0.45, primarily established using non-rotating progenitor models, suffers from significant inaccuracies when applied to rotating pre-supernova systems. The effects of metallicity and rotation on the explodability landscapes of massive stars lack robust quantification.}
   {We aim to investigate how rotation and metallicity influence the explodability of massive stars.} 
   {Using the Modules for Experiments in Stellar Astrophysics code (MESA), we simulate stars with initial rotational velocities of $V_\mathrm{ini} = 0$, $300~\mathrm{km\,s^{-1}}$, and $600~\mathrm{km\,s^{-1}}$ at three metallicities ($Z_{\odot}$, $1/10\,Z_{\odot}$, and $1/50\,Z_{\odot}$), tracking their evolution from the zero-age main sequence (ZAMS) until iron core collapse at $1000~\mathrm{km\,s^{-1}}$. For each MESA model at the onset of core collapse, we extract key parameters—enclosed mass, temperature, density, radial velocity, electron fraction, angular velocity and provide them to the 1D supernova collapse simulation code GR1D to simulate the core-collapse supernova (CCSN) phase. Through an iterative procedure, we determine the critical heating parameter $f_{\rm heat}$ within 1\% of the explosion threshold. We then define the corresponding time-averaged heating efficiency $\bar{\eta}_{\mathrm{heat}}^{\mathrm{crit}}$ at this $f_{\rm heat}$, to estimate progenitor explodability. By correlating the explosion outcomes with $\xi_{2.5}$, we derive an explodability criterion based on $\xi_{2.5}$ and also investigate the correlation between explodability and both ZAMS mass and CO-core mass across rotational velocities and metallicities.}
   {We obtain new critical values of $\xi_{2.5}$ for pre-supernova star explodability under different rotation rates and metallicities: 0.45 for models with $V_{\text{ini}}$=0; 0.48 for the $V_{\text{ini}}$=300 km s$^{-1}$ group; 0.47 for $V_{\text{ini}}$=600 km s$^{-1}$ at $Z=Z_{\odot}$, and 0.59 for low metallicity (Z=1/10 and 1/50 $Z_\odot$). These criteria enable rapid assessment of progenitor explodability for EOS configurations resembling LS220. The upper limit of pre-supernova star compactness for producing CCSNe is significantly higher in chemically homogeneous evolution (CHE) cases compared to non-CHE scenarios. This discrepancy primarily arises because the centrifugal force generated by rotational motion in pre-supernova star more effectively facilitates explosions compared to non-rotating scenarios. According to the explodability criterion of the compactness $\xi_{2.5}$, we give the ZAMS mass ranges for FSN in different models. We also determine the position of the CO-core mass corresponding to the compactness peak. Our results show  that CHE undergone by rapidly rotating low-metallicity massive stars leads to a significant decrease of the ZAMS and CO-core mass range for FSN.}
   {Rotation substantially affects the explodability of low-metallicity massive stars, underscoring the importance of incorporating rotational effects in models of core-collapse supernova progenitors.}


   \maketitle

\section{Introduction}
At the conclusion of massive star evolution, when electron degeneracy pressure is no longer sufficient to counteract gravitational forces, the iron core undergoes collapse. Upon reaching nuclear saturation compactness, neutron degeneracy pressure renders the core rigid and incompressible. The infalling material generates a shock wave as it impacts the core \citep{2017hsn..book.1095J}. However, the shock wave may weaken and stall due to two primary processes: the dissociation of heavy nuclei preceding the shock front \citep{1966ApJ...143..626C} and the subsequent loss of neutrinos, leading to the formation of an accretion shock wave \citep{2017hsn..book.1095J}. If the shock wave successfully revives and breaks out of all envelopes, the supernova will successfully explode \citep{2011ApJ...738..154H,2011MNRAS.412.1441L}, leaving a neutron star, or forming a black hole (BH) via fallback accretion \citep{2018MNRAS.477L..80K, 2018ApJ...852L..19C, 2023ApJ...957...68B, 2023arXiv230409350H}. If the shock wave revives but fails to breakout of the outer envelope, or if the shock wave fails to revive at all, the star will undergo failed supernova (FSN), forming a black hole \citep{2009A&A...499....1F,2011ApJ...730...70O,2023arXiv230409350H, 2025MNRAS.tmp..931B}. Understanding the mechanics of core-collapse supernovae (CCSNe) and predicting which progenitors will undergo explosion is of paramount importance \citep{2023MNRAS.524.4109G,2021PASA...38...62D,2009A&A...499....1F,2015PASA...32...16S,2023arXiv230409350H}.

The prevailing shock revival theory is the neutrino delayed heating mechanism proposed by \cite{1985ApJ...295...14B}, which has been broadly supported by multidimensional simulations \cite{2025arXiv250214836J}. This theory posits that hundreds of milliseconds after the initial rebound, electron-type neutrinos and antineutrinos transfer thermal energy from the  protoneutron star (PNS) to the material located behind the shock wave. Once the post-shock material attains sufficient energy, the shock wave is revitalized and continues to propagate outward \citep{2017hsn..book.1095J,1985ApJ...295...14B}. Notably, in the vast majority of one-dimensional (1D) supernova simulations, the neutrino heating mechanism proves inadequate for triggering a supernova explosion \citep{2002PhRvD..65d3001H, 2006NuPhA.777..356B, 2017PASA...34...67F} unless the neutrino heating efficiency is significantly enhanced \citep{2023MNRAS.524.4109G,2005ApJ...620..840L,2006A&A...457..281B,2017MNRAS.472..491M,2017ApJ...850...43R}. With the enhancement of computational capabilities, conducting 2D \citep{1993ApJ...415..278M,1994ApJ...435..339H} and 3D \citep{1996A&A...306..167J,2002ApJ...574L..65F} CCSNe simulations has become feasible. Multi-dimensional simulations can enable successful CCSNe explosions \citep{2015ApJ...807L..31L,2015MNRAS.453..287M,2016ApJ...818..123B,2019MNRAS.484.3307M,2020MNRAS.491.2715B,2023MNRAS.526.5900V}. The advancement of multidimensional simulations has brought to light additional explosion mechanisms, such as neutrino-driven turbulence and convection, which are increasingly recognized for their critical roles in supernova dynamics \citep{2008ApJ...688.1159M,2018ApJ...856...22M}. Consequently, assessing the potential for a supernova to explode necessitates multidimensional simulations.

However, multidimensional simulations are extremely time-consuming and computationally intensive \citep{2021ApJ...912...29B,2023MNRAS.524.4109G}. In contrast, 1D simulations are significantly faster. \cite{2018JPhG...45j4001O} conducted a comprehensive comparison of several simulation codes designed to study the CCSNe mechanism and found that the results from most codes were similar. This ensures the reliability of the code and provides a foundation for studying the effects of different parameters on supernova explodability. 

Through 1D parametric simulations, certain physical parameters of pre-supernova stars are correlated with their explosive outcomes \citep{2021ApJ...912...29B}. Utilizing these parameters allows us to make rapid assessments regarding the potential for pre-supernova stars to explode. For this, \cite{2011ApJ...730...70O} proposed the compactness parameter $\xi_{\mathrm{2.5}}$, \cite{2015ApJ...801...90P} employed the critical neutrino luminosity, \cite{2016ApJ...818..124E} also introduced a method using two parameters, and \cite{2016MNRAS.460..742M} utilized five physically motivated parameters to determine the outcomes. More recently, \cite{2025A&A...700A..20M} formulated a multi-parameter criterion based on a large set of stellar models, achieving >99\% accuracy in predicting the outcomes of a semi-analytic supernova model.

Among all the models mentioned above, the compactness parameter $\xi_{2.5}$ proposed by \cite{2011ApJ...730...70O} to predict the explodability of pre-supernova stars is the simplest and most widely applied \cite{2023arXiv230409350H}. However, some multi-dimensional simulations cast doubt on using compactness to predict pre-supernova explodability. For example: the 2D simulations of 100 stars in \cite{2023MNRAS.526.5900V} and \cite{2022MNRAS.517..543W}, along with the 3D simulations in \cite{2019MNRAS.485.3153B,2020MNRAS.491.2715B,2024ApJ...964L..16B}. The criterion $\xi_{2.5}$ is based on the 1D supernova collapse simulation code \texttt{GR1D} \citep{2010CQGra..27k4103O,2015ApJS..219...24O}. In one-dimensional simulations, due to the absence of neutrino-driven turbulence and convective motion, significant difficulties arise in evaluating explodability. Therefore, in the simulations, \cite{2011ApJ...730...70O} artificially increased the neutrino heating rate in the post-shock matter until explosions were triggered, then statistically determined the critical time-averaged heating efficiency required ($\eta_{\mathrm{heat}}^{\mathrm{crit}}$) for pre-supernova star explosion. They found that pre-supernova with high compactness ($\xi_{2.5}$) demanded extremely high $\eta_{\mathrm{heat}}^{\mathrm{crit}}$. Within the neutrino-driven paradigm, such high efficiencies were deemed unlikely to be achieved, leading to the conclusion that these progenitors would fail to explode. Ultimately, they established a predictive criterion for CCSN explodability based on the pre-supernova star compactness parameter $\xi_{2.5}$. However, this criterion was established primarily based on non-rotating pre-supernova star models. Recent studies \citep{2016MNRAS.461L.112T, 2018ApJ...852...28S, 2020MNRAS.492.4613O, 2020ApJ...901..114A} suggest that centrifugal forces in rapidly rotating massive pre-supernovae star may significantly facilitate neutrino-driven explosions. Consequently, applying the original compactness-based criterion to predict the explodability of rapidly rotating massive pre-supernova stars may lead to erroneous conclusions. 

Additionally, the evolution during the main-sequence phase can significantly influence the structure of pre-supernova stars, thereby affecting the explosion outcome \citep{2023ApJ...952...79L, 2025ApJ...979L..37L}. Extensive numerical simulations show that the final iron core mass of stars does not monotonically increase with initial mass \citep{1996ApJ...457..834T,2014ApJ...783...10S,2018ApJS..237...13L,2021A&A...645A...5S,2024A&A...682A.123T,2025A&A...695A..71L}. Changes in internal mixing, metallicity, and rotation all influence the relation between the initial and final mass of stars \citep{2020MNRAS.494L..53F}, thereby affecting explodability. Rotation can induce various instabilities during the evolutionary process \citep{2000ApJ...544.1016H}, which facilitate chemical mixing and angular momentum transfer \cite{2023arXiv230409350H}. Stellar rotation transports hydrogen from the outer layers to the core burning region, accelerating the growth of the helium core and increasing its mass. Meanwhile, rotational mixing stirs more helium into the envelope, raising the mean atomic weight of the envelope and reducing its opacity \citep{2016ApJ...821...38S,2016ARNPS..66..341J,2021MNRAS.506L..20Y}. A more massive helium core significantly enhances the stellar luminosity, while the increased helium abundance in the envelope promotes a blue solution \citep{1988Natur.334..508S,1992A&A...265L..17L}. Metallicity affects the mass loss rate, influencing the stellar mass during evolution. In particular, low metallicity combined with rapid rotation may lead to chemically homogeneous evolution(CHE) \citep{1987A&A...182..243M,1992A&A...265L..17L,2011A&A...530A.115B}, thereby enabling the pre-supernova star to retain a larger mass and rotational velocity.

Current multi-wavelength observations and theoretical studies collectively indicate that the vast majority of massive stars do not exist in isolation, but rather reside in close binary or more complex multiple stellar systems \citep{2007ApJ...670..747K, 2012Sci...337..444S, 2013A&A...550A.107S, 2014ApJS..215...15S}. Massive stars in multiple star systems exchange mass through stable Roche-lobe overflow or merger events during their evolution \citep{2022MNRAS.516.1406S, 2024A&A...682A.169H}. This implies that the majority of supernovae originate from such mass exchange objects \cite{2017ApJS..230...15M,2024A&A...686A..45S}. From pure helium star models that approximate stripped stars in binaries \citep{2022A&A...661A..60A,2023A&A...671A.134A} to binary evolution simulations that incorporate stripped stars, accretors, and mergers \citep{2021A&A...656A..58L,2021A&A...645A...5S,2023ApJ...950L...9S,2024A&A...686A..45S}, studies collectively demonstrate that binary interactions profoundly shape the pre-collapse core structures of stars. These structural changes directly impact the feasibility of neutrino-driven supernova explosions \citep{2019MNRAS.484.3307M,2021ApJ...916L...5V,2020ApJ...896...56W,2022A&A...657L...6A}. The complexity of binary/multiple stellar evolution, however, poses unique challenges that extend beyond the scope of this work. We focus on studying the impact of rotation and metallicity on the explodability of massive single stars during their evolution.

In this paper, we simulated the entire evolution of stars from the main sequence to explosion, obtaining different explodability criteria for $\xi_{2.5}$ across various model groups, and investigated the effects of rotation and metallicity on the explodability of massive stars. In Section 2, we present the model parameter settings and research methods. In Section 3, the results of the study are shown. A brief summary is provided in Section 4.

\section{Model}
In this section, we will present the model parameter settings for the Modules for Experiments in Stellar Astrophysics code (MESA) in version 10398 \citep{2011ApJS..192....3P,2013ApJS..208....4P,2015ApJS..220...15P,2018ApJS..234...34P,2019ApJS..243...10P} and GR1D.

\subsection{Pre-collapse stellar models}
We utilize the open-source 1D stellar evolution simulation code MESA to model the evolution from zero-age main sequence (ZAMS) to core collapse at 1000 km s$^{-1}$. In the simulations, we configure initial rotational velocities ($V_{\mathrm{ini}}$: 0, 300 and 600~km~s$^{-1}$) and metallicities ($Z_\odot,\ 1/10\ Z_\odot,\ 1/50\ Z_\odot$) with stellar initial masses at ZAMS spanning 10--80~M$_\odot$.

Our MESA parameter settings align with those of \cite{2023ApJ...952...79L}, with the exception of solar metallicity, which we set to 0.014 \citep{2009ARA&A..47..481A}. The mixing length parameter \texttt{$\alpha_{\mathrm{MLT}} = 1.5$} \citep{1958ZA.....46..108B}, with hydrogen burning stage overshooting \texttt{$\alpha_{\mathrm{ov}} = 0.335$} and the semiconvection mixing efficiency \texttt{$\alpha_{\mathrm{SC}} = 0.01$}. The rotation can trigger the Goldreich–Schubert–Fricke instability, Eddington–Sweet circulation, dynamical instability, and secular instability, which produce the mixing \citep{2000ApJ...528..368H}. The turbulent viscosity to diffusion coefficient ratio is $1/30$ \citep{2000ApJ...528..368H}. Stellar winds are modeled according to  \cite{2006A&A...460..199Y,2016A&A...588A..50M}: for hydrogen-rich stars ($X_{\mathrm{s}} > 0.7$), we apply the prescription by \cite{2001A&A...369..574V}; for hydrogen-poor stars ($X_{\mathrm{s}} < 0.4$), we adopt the formulas from \cite{1995A&A...299..151H}; for intermediate compositions ($0.4 \leq X_{\mathrm{s}} \leq 0.7$), linear interpolation is employed. Rotational enhancement of mass loss is included, following \cite{1993ApJ...409..429B}. Rotation is limited to $\Omega / \Omega_{\mathrm{crit}} < 0.98$ to avoid critical velocity \citep{1998A&A...329..551L,2020ApJ...901..114A}. The models are computed with the MESA \texttt{approx21} nuclear network. The parameter \texttt{mesh\_delta\_coeff} is set to 0.5, and \texttt{varcontrol\_target} is set to $3 \times 10^{-4}$.

Following \cite{2011ApJ...730...70O}, we still adopt the compactness parameter $\xi_{2.5}$ of pre-supernova stars to quantify their core compactness. The parameter $\xi_{\mathrm{M}}$ is defined as:

\begin{equation}
\xi_{\mathrm{M}} = \dfrac{M/M_{\odot}}{R\left(M_{\mathrm{bary}} = M\right) \big/ \left(1000 \, \text{km}\right)},
\label{eq:compactness}
\end{equation}
where \( M \) typically represents 2.5 $M_{\odot}$, and \( R(M) \) refers to the radius of a closed region with a baryonic mass of \( M \), which is usually calculated at the core bounce or at a collapse velocity of 1000 km s$^{-1}$. \cite{2011ApJ...730...70O} provides the criterion for the explodability: if \( \xi_{\mathrm{2.5}} < 0.45 \), a CCSN can successfully explode; otherwise, FSN may occur.

\subsection{Core-collapse model}
We simulated the CCSNe using GR1D \citep{2010CQGra..27k4103O,2015ApJS..219...24O}, an open-source, spherically symmetric, general-relativistic neutrino-radiation hydrodynamics code. The six input parameters (enclosed mass, temperature, density, radial velocity, electron fraction, and angular velocity) required for each grid point of the progenitor supernova model in GR1D are provided by MESA when the iron core collapse velocity reaches 1000 km s$^{-1}$. GR1D employs the M1 scheme \citep{2015ApJS..219...24O} for neutrino transport. The M1 transport is a sophisticated neutrino transport method that employs a two-moment scheme to solve the Boltzmann equation governing neutrino propagation \citep{2011PThPh.125.1255S,2013PhRvD..87j3004C}. This transport is more complex than the neutrino transport approach adopted in \cite{2011ApJ...730...70O}.

We employed a computational grid comprising 600 radial zones. Within a radius of 20 km, the radial spacing is set to 300 m. Beyond 20 km, the grid extends logarithmically to an outer boundary where the density reaches 2000 g cm$^{-3}$ (Higher-resolution simulations are presented in Appendix A). Our simulations include the progenitor's intrinsic rotation and incorporate an additional parametric neutrino heating term, whose magnitude is controlled by an adjustable heating coefficient. The equation of state (EOS) adopted is LS220 \citep{1991NuPhA.535..331L}, featuring a nuclear incompressibility of 220 MeV. Similarly, the neutrino opacity table was generated using NuLib. For LS220, we utilized the same NuLib table as \cite{2015ApJS..219...24O}, both utilizing three neutrino flavours and 18 energy groups. The simulations were evolved for a maximum physical time of 2 s. Remaining simulation parameters were set to the GR1D code's recommended values (broadly consistent with \cite{2015ApJS..219...24O}). This setting prevents GR1D from terminating prematurely due to constraints in the parameter configuration.

\subsection{Neutrino heating}
\begin{figure*}[htbp]
    \centering

    \begin{minipage}[b]{0.48\textwidth}
        \centering
        \includegraphics[width=\textwidth]{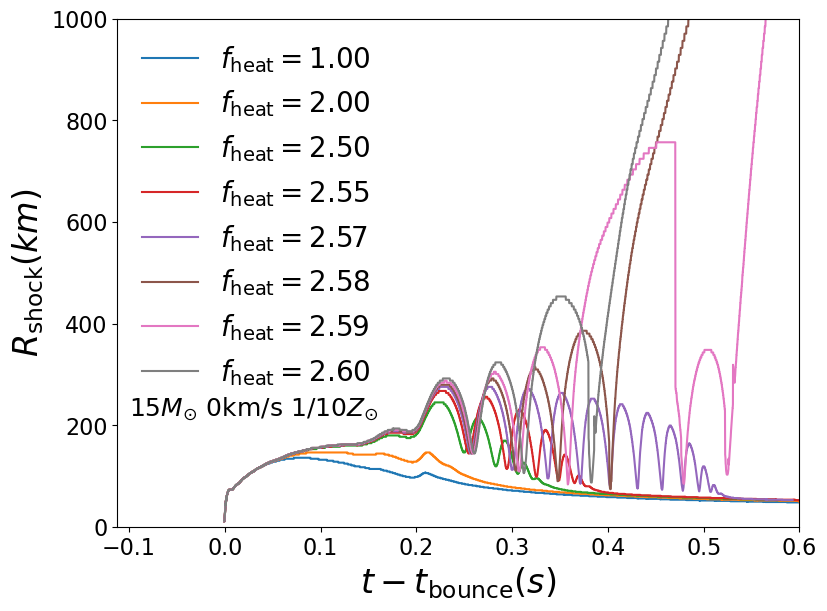}
    \end{minipage}
    \hspace{0.01\textwidth}
    \begin{minipage}[b]{0.48\textwidth}
        \centering
        \includegraphics[width=\textwidth]{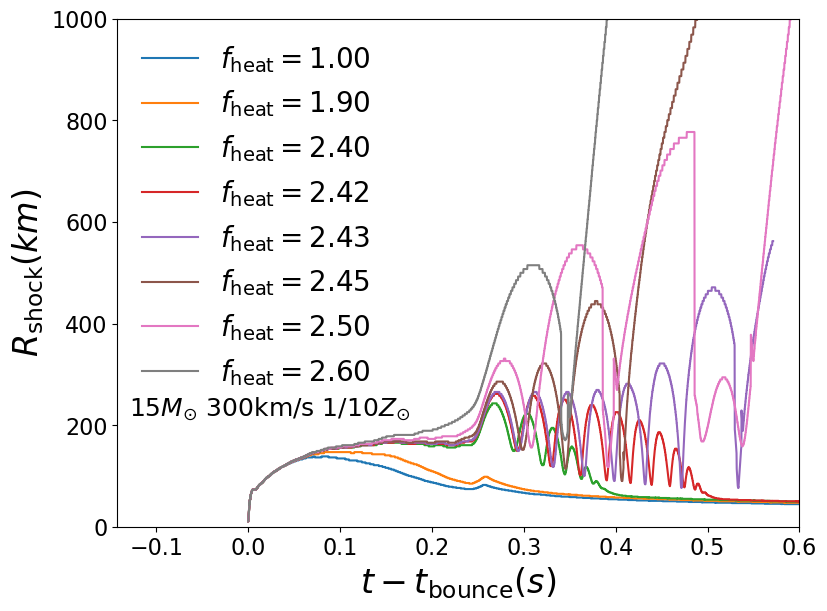}
    \end{minipage}

    \vspace{0.5cm}

    \begin{minipage}[b]{0.48\textwidth}
        \centering
        \includegraphics[width=\textwidth]{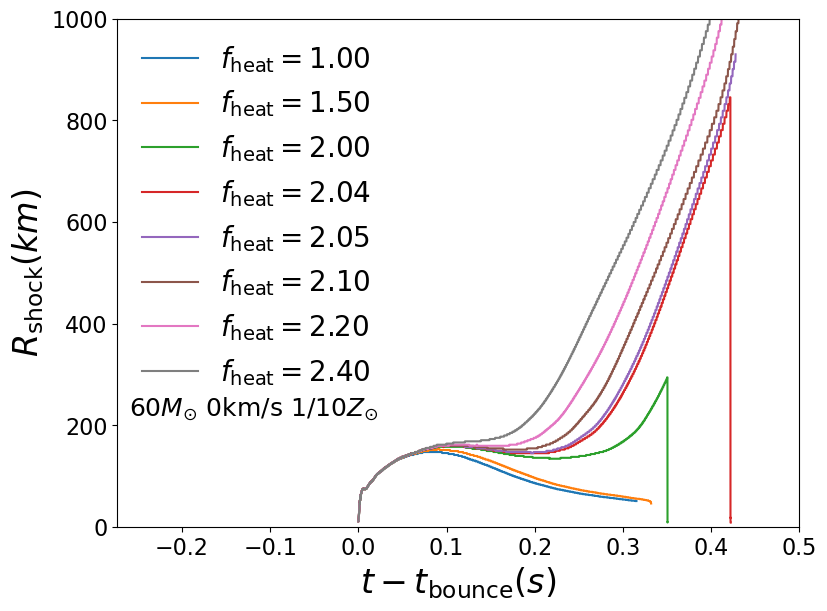}
    \end{minipage}
    \hspace{0.01\textwidth}
    \begin{minipage}[b]{0.48\textwidth}
        \centering
        \includegraphics[width=\textwidth]{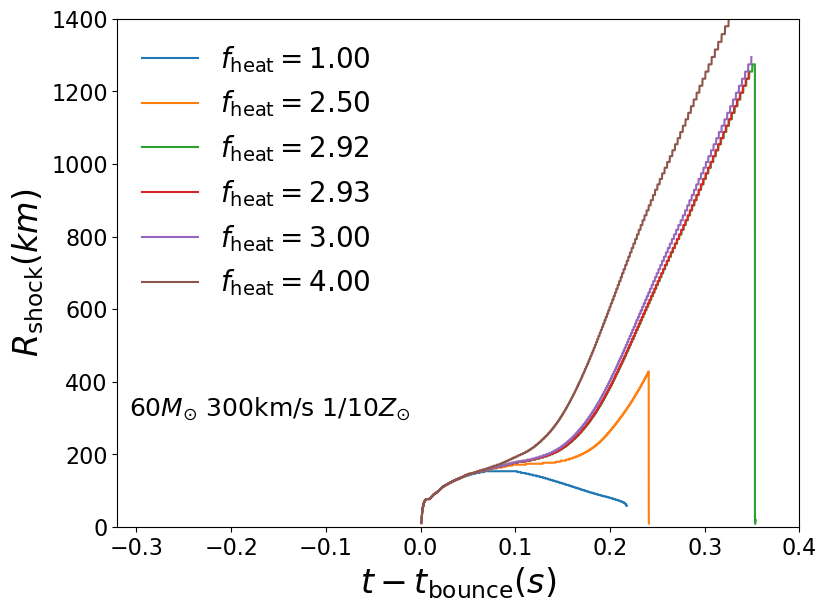}
    \end{minipage}

    \caption{The shock wave radius evolution versus explosion time $t_{\mathrm{bounce}}$ for different $f_{\mathrm{heat}}$ values. Simulations start with an iron core collapse velocity of 1000 km s$^{-1}$, where $t=0$ corresponds to shock wave formation. Top and bottom panels show 15 M$_{\odot}$ and 60 M$_{\odot}$ ZAMS masses respectively, with initial velocities $V_{\mathrm{ini}} = 0$ (left panels) and 300 km s$^{-1}$ (right panels). All models have $1/10$ Z$_{\odot}$ metallicity. }
    \label{fig:1}
\end{figure*}

The accretion history of the proto-neutron star (PNS) prior to shock revival determines the outcome: if the PNS accretes enough material to surpass its maximum supported mass (Tolman-Oppenheimer-Volkoff limit), it evolves into black hole via a FSN; otherwise, a successful CCSN explosion occurs. This approach reflects the fact that in progenitor models where neutrino heating exceeds a critical threshold, sufficient energy deposition can trigger shock revival. To achieve explosion initiation in fully self-consistent 1D spherically-symmetric simulations, artificial enhancement of neutrino energy deposition in the post-shock region is typically required. The GR1D code implements neutrino heating through the parameterization of \cite{2001A&A...368..527J}, where the heating rate at radius $r$ is:

\begin{equation}
Q^{\text{heat}}_{v_i}(r) = f_{\text{heat}} \frac{L_{v_i}(r)}{4\pi r^2} \sigma_{\text{heat},v_i} \frac{\rho}{m_{\mathrm{u}}} X_i \left\langle \frac{1}{F_{v_i}} \right\rangle e^{-2\tau_{v_i}},
\label{eq:4}
\end{equation}
where $f_{\mathrm{heat}}$ is a scaling parameter introduced to parametrically amplify the heating rate, $L_{\mathrm{\nu_i}}(r)$ is the neutrino luminosity interior to radius $r$, $i$ refers to the neutrino flavours. $\tau_{\mathrm{v_i}}$ is the optical depth determined by the leakage scheme. $\sigma_{\mathrm{heat},v_i}$ denotes the energy-averaged absorption cross section, and $X_{\mathrm{i}}$ is the corresponding mass fraction for the neutrino interactions. The term $\langle 1/F_{\mathrm{v_i}} \rangle$ represents the average inverse flux factor, which is analytically approximated as a function of the optical depth $\tau$ by comparison with the angle-dependent radiative transport calculations of \cite{2008ApJ...685.1069O}.

By increasing the $f_{\mathrm{heat}}$ value, we can enhance neutrino heating, thereby triggering CCSN explosion in 1D simulations. We adopt the same method as \citet{2011ApJ...730...70O}, iteratively adjusting $f_{\mathrm{heat}}$ to determine its critical value for explosion, $f_{\mathrm{heat}}^{\mathrm{crit}}$, within 1\% relative precision. Figure~\ref{fig:1} shows the shock radius evolution with time $t_{\mathrm{bounce}}$ for different $f_{\mathrm{heat}}$ values. Top and bottom panels show 15 M$_{\odot}$ and 60 M$_{\odot}$ ZAMS masses respectively, with initial velocities $V_{\mathrm{ini}} = 0$ (left panels) and 300 km s$^{-1}$ (right panels). All models have $1/10$ Z$_{\odot}$ metallicity. For models with varying $f_{\mathrm{heat}}$, the shock emergence times remain identical. At low $f_{\mathrm{heat}}$, no explosions occur in any of the models. As $f_{\mathrm{heat}}$ increases, the shock stagnation phase becomes prolonged. With further increases in $f_{\mathrm{heat}}$, the shock propagates outward to the stellar surface, ultimately producing a CCSN. We note that shock waves in both models of the top panel (15 M$_{\odot}$ ZAMS mass) exhibit oscillations with increasing $f_{\mathrm{heat}}$ \citep{2006A&A...447.1049B}, while remaining absent in the bottom panel (60 M$_{\odot}$). The top panel requires $f_{\mathrm{heat}} \geq 2.58$ (left) and 2.43 (right) for successful shock revival, while the bottom panel demands $f_{\mathrm{heat}} \geq 2.05$ (left) and 2.93 (right) respectively.

\subsection{Explodability criteria for CCSN}
\begin{figure*}[htbp]
    \centering

    \begin{minipage}[b]{0.48\textwidth}
        \centering
        \includegraphics[width=\textwidth]{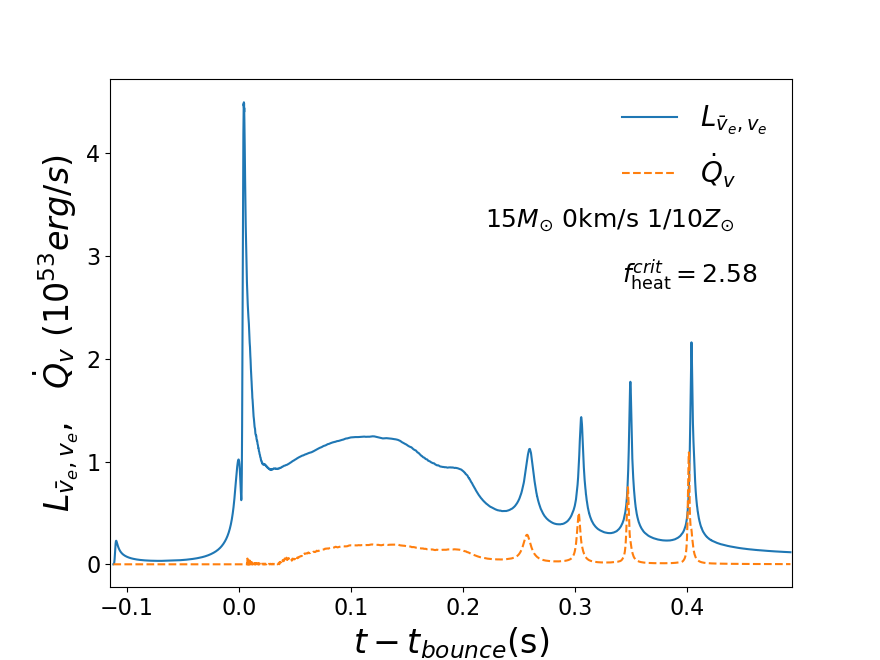}
    \end{minipage}
    \hspace{0.01\textwidth}
    \begin{minipage}[b]{0.48\textwidth}
        \centering
        \includegraphics[width=\textwidth]{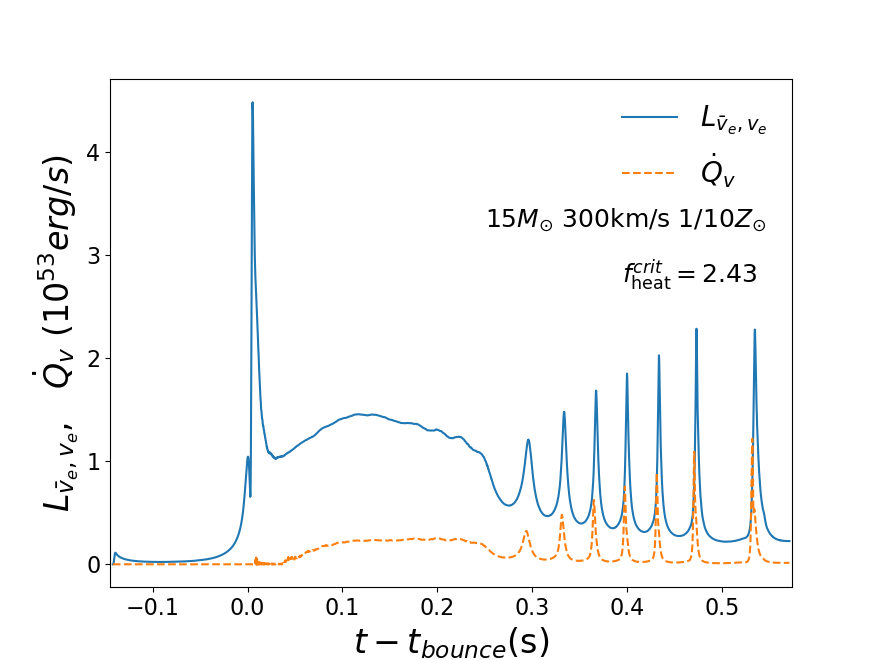}
    \end{minipage}

    \vspace{0.5cm}

    \begin{minipage}[b]{0.48\textwidth}
        \centering
        \includegraphics[width=\textwidth]{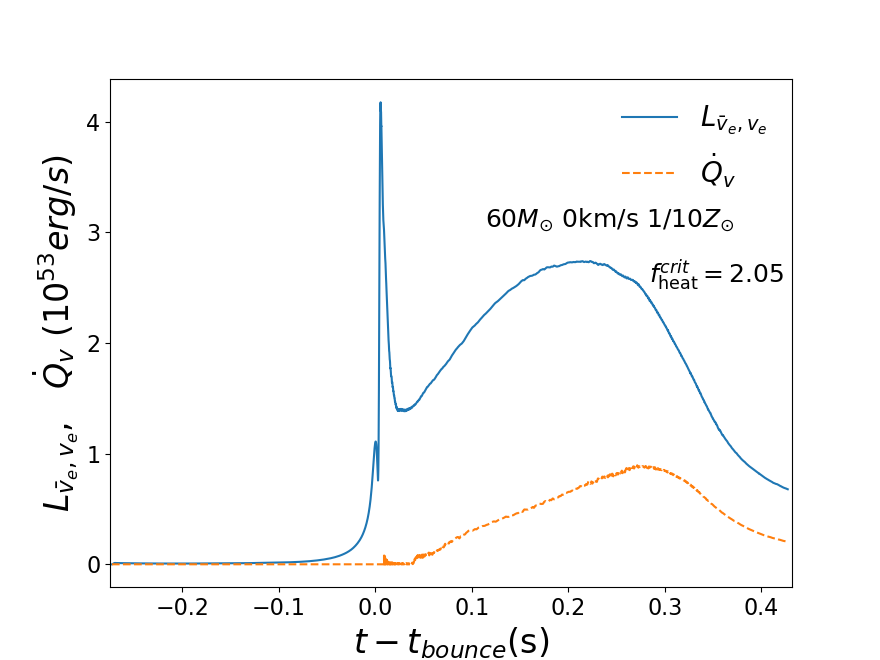}
    \end{minipage}
    \hspace{0.01\textwidth}
    \begin{minipage}[b]{0.48\textwidth}
        \centering
        \includegraphics[width=\textwidth]{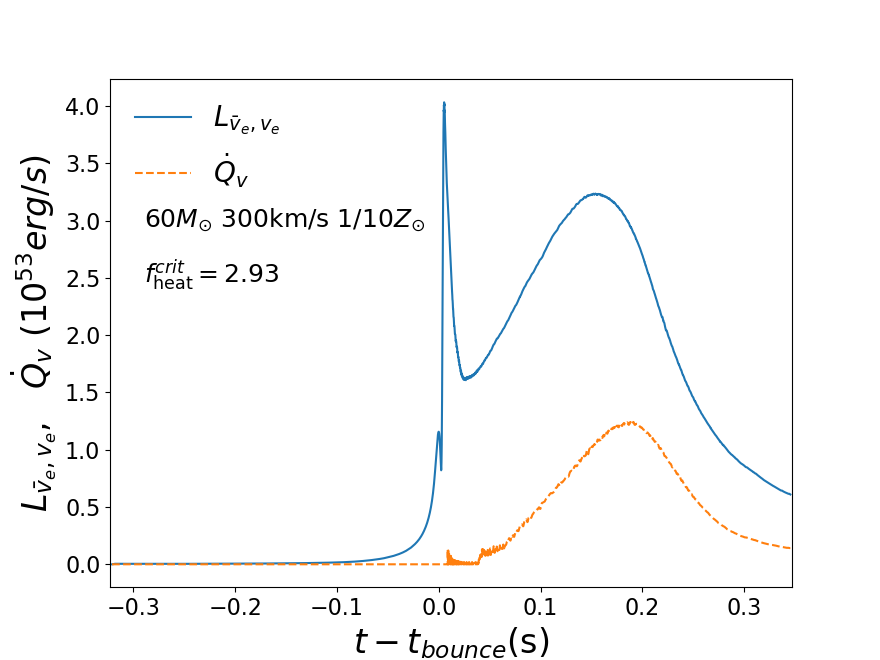}
    \end{minipage}
    \caption{For $f_{\mathrm{heat}} = f_{\mathrm{heat}}^{\mathrm{crit}}$, evolution of electron neutrino and antineutrino luminosities $L_{\bar{\nu}_e,\nu_e}$ (solid lines) at the gain radius and net neutrino energy deposition rate in the gain layer $\dot{Q}_{\nu}$ ($\int_{\text{gain}} \dot{q}^{+}_{v} \, dV$, dashed lines) versus time $t_{\mathrm{bounce}}$. Simulations start with an iron core collapse velocity of 1000 km s$^{-1}$, where $t=0$ corresponds to shock wave formation.  Top and bottom panels show 15 M$_{\odot}$ and 60 M$_{\odot}$ ZAMS masses respectively, with initial velocities $V_{\mathrm{ini}} = 0$ (left panels) and 300 km s$^{-1}$ (right panels). All models have $1/10$ Z$_{\odot}$ metallicity.}
    \label{fig:2}
\end{figure*}
While we achieve CCSN explosion in 1D simulations through artificially enhanced neutrino heating in the previous section, it remains unclear whether these models can explode under self-consistent multidimensional natural scenarios with turbulence. Therefore, \cite{2011ApJ...730...70O} used the time-averaged heating efficiency of the critical model ($f_{\mathrm{heat}}$ = $f_{\mathrm{heat}}^{\mathrm{crit}}$), $\bar{\eta}_{\mathrm{heat}}^{\mathrm{crit}}$. Here, $\bar{\eta}_{\mathrm{heat}}$ is defined as \citep{2011ApJ...730...70O}:

\begin{equation}
\bar{\eta}_{\text{heat}} = \overline{\int_{\text{gain}} \dot{q}^{+}_{v} \, dV/(L_{v_{e}} + L_{\bar{v}_{e}})_{r_{gain}}}
\label{eq:4},
\end{equation}
where \(\dot{q}^{+}_{v}\) denotes the net energy deposition rate within the gain region, and $(L_{v_{e}} + L_{\bar{v}_{e}}) _{r_{\mathrm{gain}}}$ represents the luminosity of electron neutrinos and anti-electron neutrinos at the gain radius. We calculate the average value between the shock formation time and the explosion time. The explosion time is defined as the moment when the region behind the shock attains a positive velocity and accretion onto the PNS ceases. (Note: In GR1D simulations, outward shock propagation does not necessarily imply outward motion of the material behind the shock; if such a point does not occur, we will increase $f_{\mathrm{heat}}$ to continue the simulation). $\bar{\eta}_{\mathrm{heat}}^{\mathrm{crit}}$ characterizes how much of the available neutrino energy must be redeposited on average to explode a given progenitor, independent of the transport scheme or numerical code \citep{2011ApJ...730...70O}, therefore remains equally valid for the M1 neutrino transport model. By performing 1D simulations of the confirmed eruptive pre-supernova star models, we can estimate the required range of $\bar{\eta}_{\mathrm{heat}}^{\mathrm{crit}}$ values in 1D models that correspond to successful explosions. This approach allows us to assess via 1D simulations whether a specific progenitor-EOS combination is more likely to result in explosion or BH formation.

\cite{1995ApJS..101..181W} derived $\bar{\eta}_{\mathrm{heat}}^{\mathrm{crit}} \approx 0.13$ from simulations of a 15\,$M_\odot$ ZAMS progenitor with solar metallicity. \cite{2006A&A...457..281B} artificially exploded this progenitor in 1D and found $\bar{\eta}_{\mathrm{heat}}^{\mathrm{crit}}$ ranging from 0.1 to 0.15. \cite{2011ApJ...730...70O} found that their simulated data (The EOS is LS180 and LS220) could be clearly divided into two categories: low-compactness models with $\bar{\eta}_{\mathrm{heat}}^{\mathrm{crit}} \lesssim 0.2$ (The simulations exhibit some noise, with the average value being approximately 0.16, which is close to the value of 0.15 reported in \cite{2006A&A...457..281B}) and high-compactness models with $\bar{\eta}_{\mathrm{heat}}^{\mathrm{crit}} \gtrsim 0.23$--0.27. Assuming that multi-dimensional simulations could achieve a fixed heating efficiency independent of compactness, they proposed that the former category could produce successful explosions while the latter could not. The limitation of this method lies in the fact that this assumption cannot be guaranteed; multi-dimensional simulations by \cite{2025MNRAS.tmp..931B} show that neutrino heating varies with the compactness. This work employs the $\bar{\eta}_{\mathrm{heat}}^{\mathrm{crit}}$ as the explodability criterion from \cite{2011ApJ...730...70O}, and assumes its validity for rotating pre-supernova star models.

In Figure~\ref{fig:2}, we show the temporal evolution of the electron neutrino and antineutrino luminosities $L_{\bar{\nu}_e,\nu_e}$ (solid curves) at the gain radius and the net neutrino energy deposition rate $\dot{q}^+_\nu$ (dashed curves) in the gain layer for the four models in Figure~\ref{fig:1} with $f_{\mathrm{heat}}$ = $f_{\mathrm{heat}}^{\mathrm{crit}}$. Using Equation~(3), we calculate the critical heating efficiency parameters $\bar{\eta}_{\mathrm{heat}}^{\mathrm{crit}}$, with the upper panel (left to right) corresponding to 0.112 and 0.139, and the lower panel (left to right) to 0.242 and 0.257. Based on the explodability criterion ($\bar{\eta}_{\mathrm{heat}}^{\mathrm{crit}} \lesssim 0.2$) in \cite{2011ApJ...730...70O}, we determine that the first two models (15 $M_{\odot}$) would explode, while the latter two (60 $M_{\odot}$) would form FSN. 

Please note that for the two pre-supernova star models shown in the lower panels of Figure~\ref{fig:1}, evolutionary simulations for the critical model cannot be extended further due to limitations in the EOS and neutrino opacity tables. Although shock breakout has occurred, they may lead to a FSN if the simulation can be continued, though such continuation cannot always be achieved. We demonstrate that such 'delayed failed supernova' events have limited impact on our conclusions. First, critical models  typically exhibit shock radii approaching or exceeding $1000\,\text{km}$, indicating penetration through most of the stellar envelope. Second, derived $\bar{\eta}_{\mathrm{heat}}^{\mathrm{crit}}$ values show only $\sim 3\%$ mean deviation across numerical resolutions.
\subsection{Other explosion models}
STIR model. Convection and turbulence are inherently three-dimensional phenomena. The model proposed by \cite{2020ApJ...890..127C} approximates turbulent effects in Newtonian simulations by parameterizing 3D neutrino-driven convection. \cite{2021ApJ...912...29B} extended this to a general-relativistic treatment and implemented it in the GR1D+ code. The code is computationally efficient and capable of capturing multi-D neutrino effects. Notably, at the time of writing this paper, the current implementation does not support simultaneous inclusion of STIR and rotation.

Ertl model. The 1D explosion model of \cite{2016ApJ...818..124E} and \cite{2016ApJ...821...38S} is based on the prescription by \cite{2012ApJ...757...69U}. Within their explosion model, the inner 1.1 $M_{\odot}$ core is removed after bounce and replaced by a moving inner boundary that sets both the neutrino energy and luminosity \cite{2012ApJ...757...69U}. Neutrino transport follows the gray treatment of \cite{2006A&A...457..963S}. The calibration of the explosion model required the selection of different progenitor models representing SN 1987A and the calibration of the free parameters associated with the core-boundary prescription against the observables of SN 1987A \citep{2016ApJ...818..124E, 2025MNRAS.tmp..931B}. However, \cite{2025MNRAS.tmp..931B} identify a key theoretical limitation: the model's selection of a mild contraction for the 1.1 M$_{\odot}$ shell results in neutrino average energies increasing mildly over time.

Muller model. The model in \cite{2016MNRAS.460..742M} is built upon the theoretical frameworks developed by \cite{2001A&A...368..527J}, \cite{2012ARNPS..62..407J}, and \cite{2015MNRAS.453..287M}, among other studies, and expands upon and refines these frameworks. It does not require complex hydrodynamic simulations but instead predicts the properties of neutrino-driven explosions based on the pre-collapse stellar structure. The Muller model incorporates the effects of neutrino-driven convection by substituting them with the shock radius multiplied by a scaling factor $\alpha_{turb}$. This leads to the convection in the explosion model exerting roughly the same influence on all progenitors. However, multi-D simulations reveal that convection strength correlates strongly with compactness \citep{2025MNRAS.tmp..931B}.

These 1D explosion models each have inherent limitations. We chose the explosion model from \cite{2011ApJ...730...70O} because the compactness parameter $\xi_{2.5}$ is a widely adopted and simplest parameter for predicting the explodability of pre-supernovae star, and GR1D is an open-source software.

\section{Results}

In this chapter, we compute the critical heating efficiency parameter $\bar{\eta}_{\mathrm{heat}}^{\mathrm{crit}}$ for each model to assess whether these models can undergo supernova explosions under natural conditions where the nuclear EOS similar to LS220. The explosion outcomes are then correlated with the compactness parameter $\xi_{2.5}$. Therefore, according to $\bar{\eta}_{\mathrm{heat}}^{\mathrm{crit}}=0.2$, we can establish the explodability criteria of $\xi_{\rm 2.5}$ (It can be calculated directly by MESA ) for the models with different metallicities and rotational velocities. By analyzing the influence of rapid rotation and metallicity on the explodability of massive stars, we can link their explosion outcomes with their ZAMS initial masses.  
\subsection{Explodability criteria of $\xi_{\rm 2.5}$}
\begin{table*}[http]
\centering
\scriptsize 
\setlength{\tabcolsep}{3pt} 
\begin{minipage}[t]{0.49\textwidth}
\centering
\begin{tabular}{@{}ccccccccc@{}} 
\toprule
M ($M_{\odot}$) & $\xi_{2.5}$ & $M_{bary} \, (M_{\odot})$ & $M_{grav} \, (M_{\odot})$ & $t_{bounce} \, (s)$& $t_{\mathrm{bounce}} \, (s)$ & $f^{crit}_{heat}$ & $\bar{\eta}_{\text{heat}}^{\text{crit}}$ & $Explosion$ \\
\midrule
\multicolumn{9}{c}{$V_{ini}$ = 0 \text{km s$^{-1}$} $\; \; \; \; \;Z_{\odot}$} \\
\midrule
14 & 0.164 & 1.82 & 1.75 & 0.145 & 0.723 & 2.47 & 0.152 & yes \\
15 & 0.216 & 1.93 & 1.84 & 0.216 & 0.662 & 2.28 & 0.185 & yes \\
19 & 0.441 & 2.26 & 2.13 & 0.232 & 0.733 & 1.98 & 0.198 & yes \\
25 & 0.140 & 1.70 & 1.64 & 0.109 & 0.634 & 2.64 & 0.139 & yes \\
30 & 0.495 & 2.35 & 2.20 & 0.217 & 0.877 & 1.94 & 0.227 & no  \\
33 & 0.675 & 2.38 & 2.25 & 0.256 & 0.846 & 1.93 & 0.266 & no  \\
45 & 0.404 & 2.06 & 1.96 & 0.182 & 0.623 & 2.13 & 0.162 & yes \\
55 & 0.558 & 2.37 & 2.23 & 0.250 & 0.945 & 1.93 & 0.239 & no  \\
80 & 0.910 & 2.25 & 2.17 & 0.345 & 0.562 & 4.75 & 0.247 & no \\
\midrule
\multicolumn{9}{c}{$V_{ini}$ = 0 \text{km s$^{-1}$} $\; \; \; \; \;1/10\,Z_{\odot}$} \\
\midrule
14 & 0.201 & 1.91 & 1.83 & 0.149 & 0.753 & 2.29 & 0.189 & yes \\
15 & 0.120 & 1.91 & 1.83 & 0.106 & 0.620 & 2.58 & 0.112 & yes \\
17 & 0.287 & 2.02 & 1.93 & 0.184 & 0.787 & 2.20 & 0.188 & yes \\
21 & 0.466 & 2.30 & 2.16 & 0.196 & 0.794 & 1.96 & 0.221 & no  \\
23 & 0.438 & 2.23 & 2.10 & 0.171 & 0.787 & 1.98 & 0.188 & yes \\
29 & 0.321 & 2.18 & 2.06 & 0.150 & 0.655 & 2.03 & 0.193 & yes \\
32 & 0.539 & 2.33 & 2.19 & 0.217 & 0.754 & 1.92 & 0.227 & no  \\
35 & 0.594 & 2.34 & 2.21 & 0.250 & 0.863 & 1.90 & 0.251 & no  \\
45 & 0.398 & 2.04 & 1.95 & 0.205 & 0.675 & 2.21 & 0.173 & yes \\
60 & 0.782 & 2.41 & 2.30 & 0.273 & 0.701 & 2.05 & 0.242 & no  \\
80 & 0.877 & 2.31 & 2.22 & 0.266 & 0.513 & 3.59 & 0.239 & no  \\
\midrule
\multicolumn{9}{c}{$V_{ini}$ = 0 \text{km s$^{-1}$} $\; \; \; \; \;1/50\,Z_{\odot}$} \\
\midrule
13 & 0.159 & 1.82 & 1.75 & 0.094 & 0.600 & 2.45 & 0.158 & yes \\
15 & 0.165 & 1.79 & 1.72 & 0.165 & 0.690 & 2.50 & 0.135 & yes \\
17 & 0.312 & 2.17 & 2.04 & 0.161 & 0.728 & 2.08 & 0.197 & yes \\
23 & 0.442 & 2.24 & 2.12 & 0.188 & 0.683 & 1.98 & 0.190 & yes \\
29 & 0.393 & 2.21 & 2.09 & 0.170 & 0.652 & 2.00 & 0.190 & yes \\
30 & 0.427 & 2.28 & 2.14 & 0.121 & 0.631 & 1.96 & 0.197 & yes \\
31 & 0.480 & 2.32 & 2.19 & 0.131 & 0.749 & 1.92 & 0.233 & no  \\
33 & 0.587 & 2.34 & 2.21 & 0.209 & 0.826 & 1.90 & 0.250 & no  \\
45 & 0.405 & 2.17 & 2.05 & 0.202 & 0.747 & 2.09 & 0.188 & yes \\
50 & 0.500 & 2.31 & 2.17 & 0.226 & 0.987 & 1.96 & 0.237 & no  \\
60 & 0.810 & 2.42 & 2.30 & 0.307 & 0.709 & 2.18 & 0.248 & no  \\
80 & 0.871 & 2.20 & 2.13 & 0.300 & 0.520 & 2.43 & 0.222 & no  \\
\midrule
\multicolumn{9}{c}{$V_{ini}$ = 300 \text{km s$^{-1}$} $\; \; \; \; \;Z_{\odot}$} \\
\midrule
15 & 0.199 & 1.88 & 1.81 & 0.158 & 0.732 & 2.40 & 0.153 & yes \\
17 & 0.339 & 2.12 & 2.01 & 0.200 & 0.721 & 2.10 & 0.188 & yes \\
19 & 0.408 & 2.30 & 2.16 & 0.212 & 0.707 & 1.99 & 0.180 & yes \\
20 & 0.620 & 2.37 & 2.25 & 0.239 & 0.747 & 1.93 & 0.224 & no  \\
27 & 0.552 & 2.36 & 2.22 & 0.247 & 0.815 & 1.92 & 0.226 & no  \\
32 & 0.649 & 2.35 & 2.20 & 0.210 & 0.852 & 1.88 & 0.263 & no  \\
40 & 0.437 & 2.18 & 2.04 & 0.164 & 0.927 & 2.05 & 0.171 & yes \\
50 & 0.484 & 2.32 & 2.18 & 0.219 & 0.847 & 1.98 & 0.195 & yes \\
60 & 0.523 & 2.33 & 2.20 & 0.231 & 0.863 & 1.95 & 0.223 & no  \\
80 & 0.906 & 2.40 & 2.34 & 0.345 & 0.646 & 2.88 & 0.256 & no  \\
\midrule
\multicolumn{9}{c}{$V_{ini}$ = 300 \text{km s$^{-1}$} $\; \; \; \; \;1/10\,Z_{\odot}$} \\
\midrule
13 & 0.205 & 1.96 & 1.88 & 0.055 & 0.521 & 2.28 & 0.178 & yes \\
15 & 0.171 & 1.84 & 1.77 & 0.142 & 0.713 & 2.43 & 0.139 & yes \\
16 & 0.494 & 2.29 & 2.16 & 0.230 & 0.822 & 1.95 & 0.226 & no \\
21 & 0.452 & 2.26 & 2.14 & 0.142 & 0.652 & 1.96 & 0.179 & yes \\
\bottomrule
\end{tabular}
\end{minipage}
\hfill
\begin{minipage}[t]{0.49\textwidth}
\centering
\begin{tabular}{@{}ccccccccc@{}}
\toprule
M ($M_{\odot}$) & $\xi_{2.5}$ & $M_{bary} \, (M_{\odot})$ & $M_{grav} \, (M_{\odot})$ & $t_{bounce} \, (s)$& $t_{\mathrm{bounce}} \, (s)$ & $f^{crit}_{heat}$ & $\bar{\eta}_{\text{heat}}^{\text{crit}}$ & $Explosion$ \\
\hline
26 & 0.291 & 2.12 & 2.02 & 0.131 & 0.651 & 2.10 & 0.169 & yes \\
27 & 0.397 & 2.19 & 2.08 & 0.117 & 0.614 & 2.00 & 0.198 & yes \\
28 & 0.537 & 2.26 & 2.15 & 0.237 & 0.728 & 1.96 & 0.224 & no \\
35 & 0.593 & 2.30 & 2.16 & 0.184 & 0.784 & 1.90 & 0.227 & no \\
60 & 0.942 & 2.46 & 2.36 & 0.321 & 0.665 & 2.93 & 0.257 & no \\
\midrule
\multicolumn{9}{c}{$V_{ini}$ = 300 \text{km s$^{-1}$} $\; \; \; \; \;1/50\,Z_{\odot}$} \\
\midrule
15 & 0.205 & 1.88 & 1.81 & 0.151 & 0.723 & 2.40 & 0.151 & yes \\
16 & 0.340 & 2.15 & 2.03 & 0.179 & 0.602 & 2.10 & 0.168 & yes \\
19 & 0.316 & 2.08 & 1.98 & 0.188 & 0.617 & 2.15 & 0.172 & yes \\
20 & 0.513 & 2.32 & 2.20 & 0.226 & 0.760 & 1.96 & 0.227 & no \\
21 & 0.422 & 2.26 & 2.13 & 0.202 & 0.748 & 2.00 & 0.192 & yes \\
26 & 0.366 & 2.19 & 2.07 & 0.156 & 0.592 & 2.01 & 0.177 & yes \\
27 & 0.473 & 2.26 & 2.13 & 0.194 & 0.621 & 1.96 & 0.183 & yes \\
28 & 0.590 & 2.38 & 2.24 & 0.172 & 0.713 & 1.90 & 0.229 & no \\
30 & 0.656 & 2.36 & 2.22 & 0.236 & 0.773 & 1.92 & 0.252 & no \\
35 & 0.664 & 2.33 & 2.22 & 0.249 & 0.823 & 2.06 & 0.222 & no \\
40 & 0.803 & 2.62 & 2.49 & 0.249 & 0.857 & 4.49 & 0.251 & no \\
\midrule
\multicolumn{9}{c}{$V_{ini}$ = 600 \text{km s$^{-1}$}$ \; \; \; \; \;Z_{\odot}$} \\
\midrule
15 & 0.237 & 1.94 & 1.86 & 0.186 & 0.767 & 2.34 & 0.177 & yes \\
21 & 0.470 & 1.96 & 2.05 & 0.194 & 0.663 & 2.15 & 0.191 & yes \\
22 & 0.539 & 2.21 & 2.09 & 0.200 & 0.684 & 2.02 & 0.190 & yes \\
23 & 0.574 & 2.29 & 2.16 & 0.238 & 0.742 & 1.97 & 0.198 & yes \\
32 & 0.474 & 2.35 & 2.20 & 0.177 & 0.885 & 1.88 & 0.236 & no  \\
34 & 0.395 & 2.32 & 2.19 & 0.187 & 0.912 & 1.89 & 0.223 & no  \\
40 & 0.448 & 2.30 & 2.14 & 0.175 & 0.741 & 1.98 & 0.222 & no  \\
50 & 0.611 & 2.41 & 2.23 & 0.270 & 0.751 & 1.97 & 0.234 & no  \\
60 & 0.817 & 2.42 & 2.30 & 0.285 & 0.753 & 2.10 & 0.265 & no  \\
80 & 0.870 & 2.39 & 2.26 & 0.304 & 0.532 & 4.86 & 0.243 & no \\
\midrule
\multicolumn{9}{c}{$V_{ini}$ = 600 \text{km s$^{-1}$}$ \; \; \; \; \;1/10\,Z_{\odot}$} \\
\midrule
12 & 0.563 & 2.35 & 2.22 & 0.242 & 0.854 & 1.99 & 0.230 & no \\
13 & 0.432 & 2.20 & 2.09 & 0.207 & 0.732 & 2.08 & 0.199 & yes \\
15 & 0.145 & 1.75 & 1.69 & 0.145 & 0.692 & 2.75 & 0.148 & yes \\
20 & 0.566 & 2.35 & 2.23 & 0.209 & 0.639 & 2.17 & 0.192 & yes \\
21 & 0.625 & 2.29 & 2.19 & 0.156 & 0.608 & 2.43 & 0.234 & no \\
22 & 0.656 & 2.31 & 2.20 & 0.162 & 0.682 & 2.36 & 0.237 & no \\
28 & 0.519 & 2.33 & 2.18 & 0.151 & 0.809 & 1.99 & 0.183 & yes \\
40 & 0.537 & 2.42 & 2.29 & 0.270 & 0.737 & 2.22 & 0.154 & yes \\
45 & 0.745 & 2.49 & 2.37 & 0.243 & 0.703 & 2.12 & 0.243 & no \\
55 & 0.885 & 2.46 & 2.35 & 0.253 & 0.672 & 2.53 & 0.263 & no \\
\midrule
\multicolumn{9}{c}{$V_{ini}$ = 600 \text{km s$^{-1}$}$ \; \; \; \; \;1/50\,Z_{\odot}$} \\
\midrule
15 & 0.511 & 2.36 & 2.23 & 0.219 & 0.738 & 2.14 & 0.191 & yes \\
16 & 0.576 & 2.40 & 2.27 & 0.235 & 0.734 & 2.10 & 0.185 & yes \\
19 & 0.640 & 2.46 & 2.31 & 0.205 & 0.673 & 2.42 & 0.242 & no \\
22 & 0.440 & 2.16 & 2.04 & 0.157 & 0.744 & 2.08 & 0.194 & yes \\
27 & 0.365 & 2.30 & 2.14 & 0.183 & 0.852 & 2.38 & 0.118 & yes \\
28 & 0.413 & 2.33 & 2.18 & 0.192 & 1.240 & 2.29 & 0.168 & yes \\
29 & 0.432 & 2.35 & 2.20 & 0.201 & 1.150 & 2.28 & 0.153 & yes \\
31 & 0.589 & 2.46 & 2.33 & 0.249 & 0.907 & 2.40 & 0.192 & yes \\
33 & 0.645 & 2.51 & 2.38 & 0.262 & 0.820 & 2.32 & 0.238 & no \\
40 & 0.556 & 2.35 & 2.20 & 0.217 & 0.953 & 1.90 & 0.196 & yes \\
50 & 0.802 & 2.48 & 2.36 & 0.309 & 0.770 & 2.35 & 0.258 & no \\
60 & 0.874 & 2.49 & 2.37 & 0.302 & 0.891 & 3.99 & 0.283 & no \\
\bottomrule
\end{tabular}
\end{minipage}
\caption{ZAMS mass, $\xi_{2.5}$, $M_{\mathrm{bary}}$ and $M_{\mathrm{grav}}$ are the baryonic and gravitational masses, enclosed within the radius where the density equals $10^{12}~\mathrm{g\,cm^{-3}}$, evaluated at the end of simulations with $f_{\rm heat} = f_{\rm heat}^{\rm crit}$., shock formation time ($t_{\mathrm{bounce}}$), explosion time ($t_{\mathrm{f}}$), critical heating parameter at explosion ($f_{\mathrm{heat}}^{\mathrm{crit}}$), and corresponding ($\bar{\eta}_{\mathrm{heat}}^{\mathrm{crit}}$)}
\label{tab:collapsed}
\end{table*}

\begin{figure*}[htbp]
    \centering
    \begin{minipage}[b]{\textwidth}
        \centering
        \includegraphics[width=\textwidth]{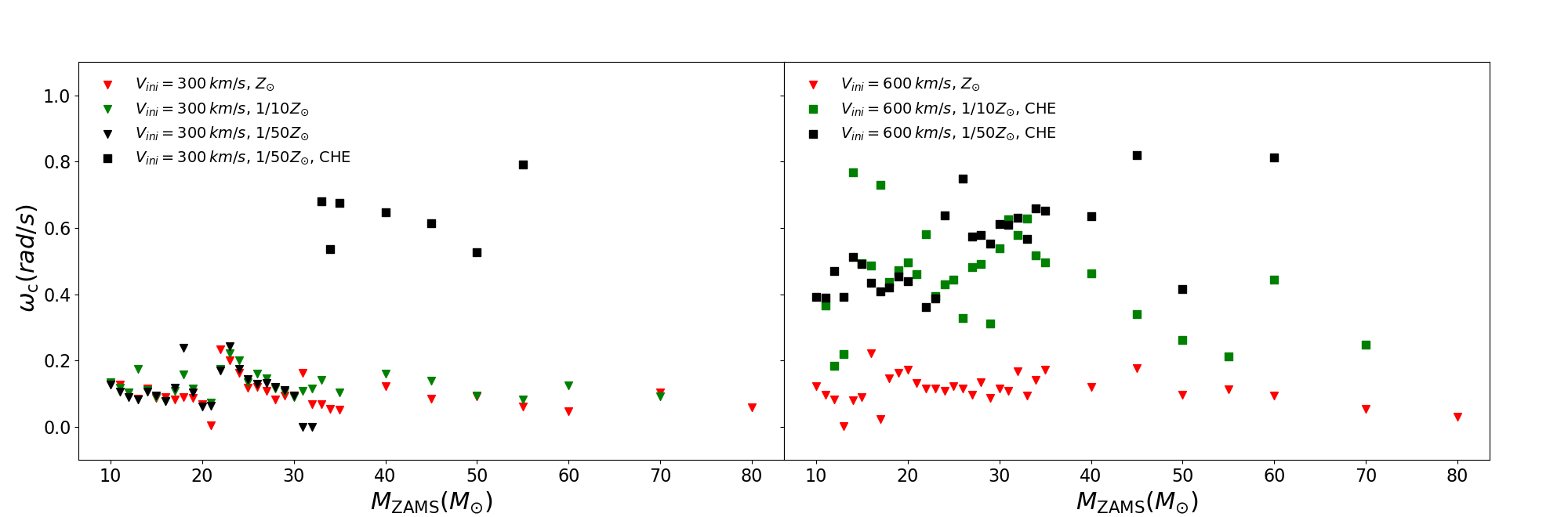}
    \end{minipage}
    \label{fig:fourimages}
    \caption{The central angular velocity ($\omega_{c}$) at pre-supernova star for different ZAMS mass models versus various $V_{\mathrm{ini}}$ and metallicities. The left and right panels represent $V_{\mathrm{ini}}$ = 300 km s$^{-1}$ and $V_{\mathrm{ini}}$ = 600 km s$^{-1}$, respectively. Red, green, and black represent metallicities of $Z_{\odot}$, $1/10$ $Z_{\odot}$, and $1/50$ $Z_{\odot}$, respectively. Squares mark the occurrence of CHE.}
    \label{fig:3} 
\end{figure*}
Due to limited computational resources, we selected only a subset of stars from each group for GR1D simulations, preferentially choosing stars with $ \xi_{\mathrm{2.5}}$ values closest to the threshold.

Figure~\ref{fig:3} shows the central angular velocity versus $M_{ZAMS}$ in pre-supernova star models with $V_{\rm ini}$ = 300 km s$^{-1}$ (left) and 600 km s$^{-1}$ (right), where different colors indicate distinct metallicities. CHE occurs in all pre-supernova star models with $V_{\mathrm{ini}}=600\ \mathrm{km\ s^{-1}}$ at both $1/10$ $Z_{\odot}$ and $1/50$ $Z_{\odot}$ metallicities. For the $V_{\mathrm{ini}}=300\ \mathrm{km\ s^{-1}}$ and $1/50$ $Z_{\odot}$ subset, CHE is exclusively achieved by stars with initial masses $M_{\mathrm{ZAMS}} \geq 33M_{\odot}$. These CHE progenitors exhibit systematically higher central angular velocities ($\omega_{\mathrm{core}}$), predominantly in the $0.3$--$0.7\ \mathrm{rad\ s^{-1}}$ range. In contrast, Non-CHE progenitors exhibit significantly lower central angular velocities ($\omega_{\mathrm{core}}$ $<$ 0.2 rad s$^{-1}$), which may affect the compactness required for the formation of FSN.

\begin{figure*}[htbp]
    \centering
    \begin{minipage}[b]{\textwidth}
        \centering
        \includegraphics[width=\textwidth]{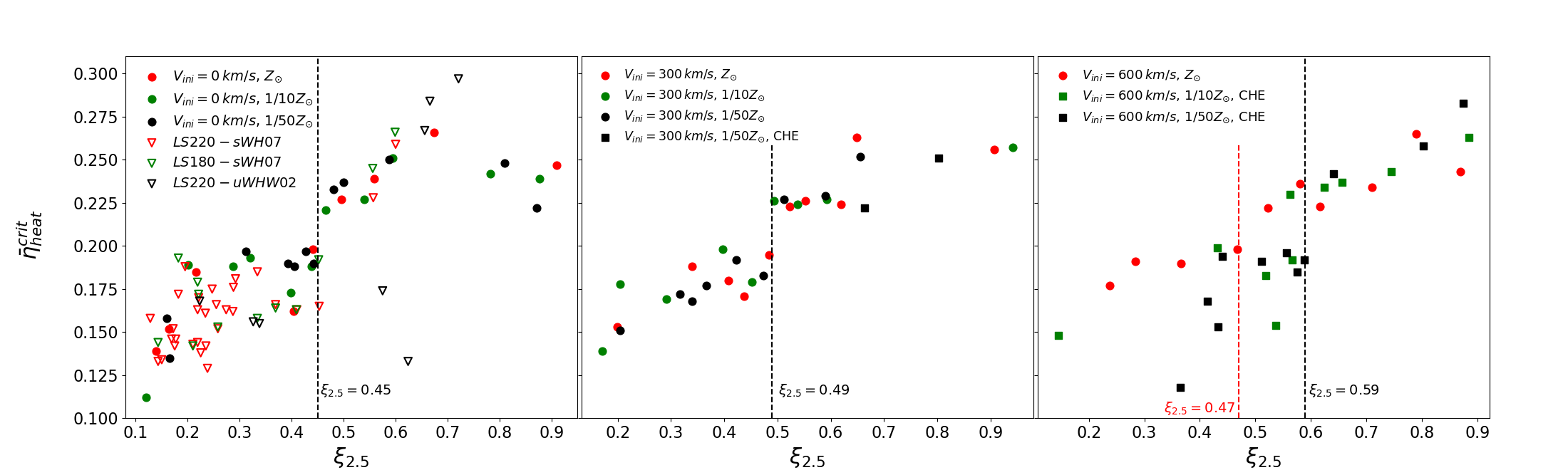}
    \end{minipage}
    \label{fig:fourimages}
    \caption{The compactness parameter $\xi_{\mathrm{2.5}}$ versus the critical model's time-averaged heating efficiency $\bar{\eta}_{\mathrm{heat}}^{\mathrm{crit}}$. Panels left, center, and right correspond to initial velocities $V_{\mathrm{ini}}$ = 0, 300 km s$^{-1}$, and 600 km s$^{-1}$, respectively. Data points in red, green, and black denote metallicities $Z_{\odot}$, 1/10 $Z_{\odot}$, and 1/50 $Z_{\odot}$. Squares mark the occurrence of CHE. The hollow inverted triangles are calculated by \cite{2011ApJ...730...70O}, with red, green, and black colors representing the combinations of EOS and models: LS220-sWH07, LS180-sWH07, and LS220-uWHW02, respectively.}
    \label{fig:4}
\end{figure*}

Figure~\ref{fig:4} shows the $\xi_{2.5}$ values versus their corresponding $\bar{\eta}_{\mathrm{heat}}^{\mathrm{crit}}$ for all stellar models, with the complete dataset tabulated in Table~1. It can be observed that $\bar{\eta}_{\text{heat}}^{\text{crit}}$ increases with the augmentation of $\xi_{2.5}$. Obviously, all models in this work can be divided into two parts well by $\bar{\eta}_{\mathrm{heat}}^{\mathrm{crit}}$: the lower left zone with $\bar{\eta}_{\mathrm{heat}}^{\mathrm{crit}}\leq 0.2$, and the upper right zone with $\bar{\eta}_{\mathrm{heat}}^{\mathrm{crit}} \gtrsim 0.23$. Based on the explodability criterion of \cite{2011ApJ...730...70O}, the former would successfully undergo CCSN, while the latter would evolve into FSN.

The left panel represents models with ZAMS rotation of zero. The hollow inverted triangular data points are from \cite{2011ApJ...730...70O}, which include both LS180 and LS220 EOS with metallicities of $Z_{\odot}$ and $10^{-4}Z_{\odot}$. A unified criterion of $\xi_{2.5} = 0.45$ is established for all non-rotating models across three metallicities in our work, consistent with the results of \cite{2011ApJ...730...70O}. The two models (lower-right in left panel) from \cite{2011ApJ...730...70O} exhibit high compactness ($\xi_{2.5}>0.45$) but still explode with $\bar{\eta}_{\mathrm{heat}}^{\mathrm{crit}} < 0.2$. This is because these models feature compositional interfaces where the density drops by 50\%; when the shock reaches this interface, it rapidly breaks out, leading to CCSN explosion \citep{2011ApJ...730...70O}.

For the middle panel with $V_{\mathrm{ini}} = 300~\mathrm{km~s^{-1}}$, the majority of pre-supernova star did not undergo CHE, retaining low rotation rates ($\omega_{c} \leq 0.2$; Fig~\ref{fig:3} (left panel)). Stars with 1/50 $Z_{\odot}$ and $M_{\text{ZAMS}} \geq 33 M_{\odot}$ undergo CHE; however, none of our selected models explode because of their exceptionally high compactness. We still attempt to construct a unified explodability criterion across metallicities and derive $\xi_{2.5} = 0.48$.

The right panel, with $V_{\mathrm{ini}} = 600~\mathrm{km~s^{-1}}$, $Z_{\odot}$ models exhibit lower pre-supernova star rotational velocities due to their enhanced mass-loss rates (Fig~\ref{fig:3} (right panel)), while the $1/10\,Z_{\odot}$ and $1/50\,Z_{\odot}$ models maintain higher rotation before collapse as a result of CHE. The $\xi_{2.5}$ ranges for CCSN explosions between these two groups exhibit significant differences. No unified criterion can be established: we obtain $\xi_{2.5} = 0.47$ for non-CHE models and $\xi_{2.5} = 0.59$ for CHE models. Using the pre-supernova star models simulated by \cite{2020ApJ...901..114A} ($V_{\rm ini}$ = 600 km s$^{-1}$, Z = 1/50 $Z_{\odot}$), \cite{2023ApJ...944L..38H} calculated the compactness parameter $\xi_{2.5}$ for rapidly rotating massive stars, and they derived $\xi_{\mathrm{2.5}} \sim 0.6$ for the explodability criterion, which is consistent with our results. However, the $12\,M_{\odot}$ star with $1/10\,Z_{\odot}$ can not successfully explode although $\xi_{2.5} < 0.59$ as $\bar{\eta}_{\mathrm{heat}}^{\mathrm{crit}} > 0.22$. This is primarily due to its progenitor's low central rotation speed (weak centrifugal support) combined with excessive compactness.

For different model groups, we obtain distinct critical compactness criteria for explodability. By comparing with the angular velocity $\omega_{\mathrm{c}}$ at the pre-supernova star center in figure~\ref{fig:3}, we attribute this difference primarily to rotation. Higher pre-supernova star rotational velocities generate stronger centrifugal forces, which facilitate the explosion \citep{2016MNRAS.461L.112T, 2018ApJ...852...28S, 2020ApJ...901..114A, 2020MNRAS.492.4613O} and thereby increase the upper limit of the critical compactness required for explosion. 

Single parameters cannot always reliably predict progenitor explosibility. Furthermore, the evolution of stars on the main sequence is inherently complex: even for ZAMS stars with identical rotation and metallicity, their pre-supernova stars may exhibit significantly different density profiles and rotational properties. These factors collectively make establishing a unified explodability criterion for model groups with identical ZAMS conditions but varying masses a challenging task. Our results reflect an overarching trend across all parameters.

\subsection{Explodability of massive stars based on ZAMS and CO-core masses}
\begin{figure*}[htbp]
    \centering
    \begin{minipage}[b]{\textwidth}
        \centering
        \includegraphics[width=\textwidth]{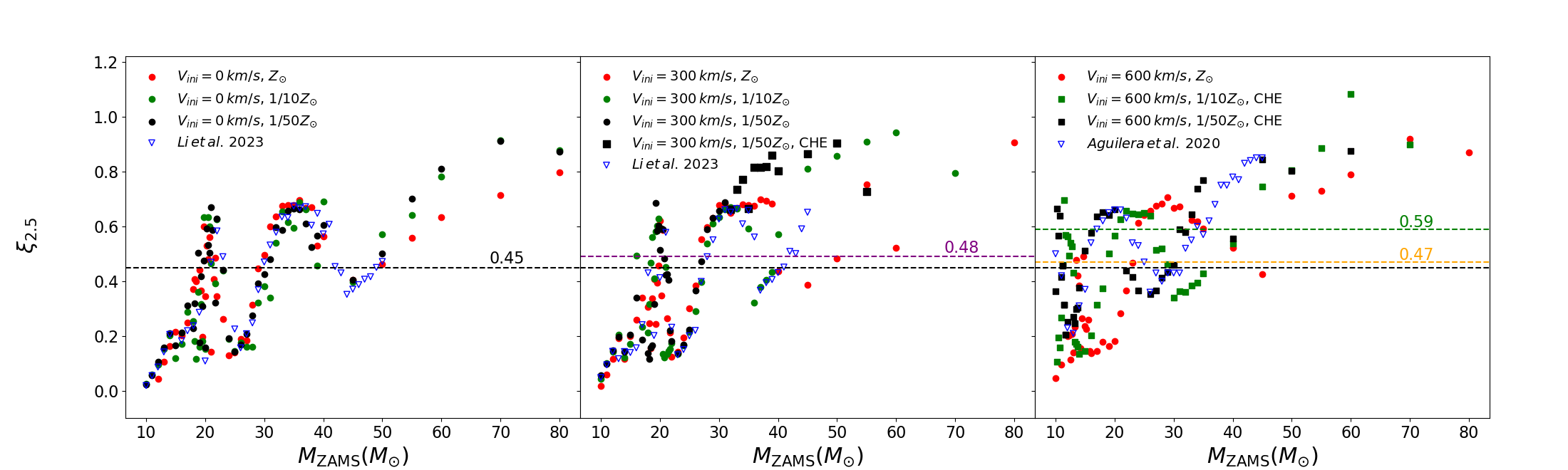}
    \end{minipage}
    \label{fig:fourimages}
    \caption{The $\xi_{\mathrm{2.5}}$ versus initial mass for models with varying initial rotational velocities $V_{\rm ini}$ and metallicities. The left, middle, and right panels correspond to models with $V_{\mathrm{ini}} = 0$, $300\,\mathrm{km\,s^{-1}}$, and $600\,\mathrm{km\,s^{-1}}$, respectively. Red, green, and black data points represent metallicities of $Z_{\odot}$, $1/10\,Z_{\odot}$, and $1/50\,Z_{\odot}$, respectively. Squares mark the occurrence of CHE. The left and middle blue triangles indicate model calculations from \cite{2023ApJ...952...79L} ($Z = 0.0017$), while the right blue triangle denotes the rapidly rotating model from \cite{2020ApJ...901..114A} ($1/50\,Z_{\odot}$). Dashed lines mark explosion criteria for $\xi_{2.5}$: the black line from \cite{2011ApJ...730...70O}, and others from our results.}
    \label{fig:5}
\end{figure*}

\begin{figure*}[htbp]
    \centering
    \begin{minipage}[b]{\textwidth}
        \centering
        \includegraphics[width=\textwidth]{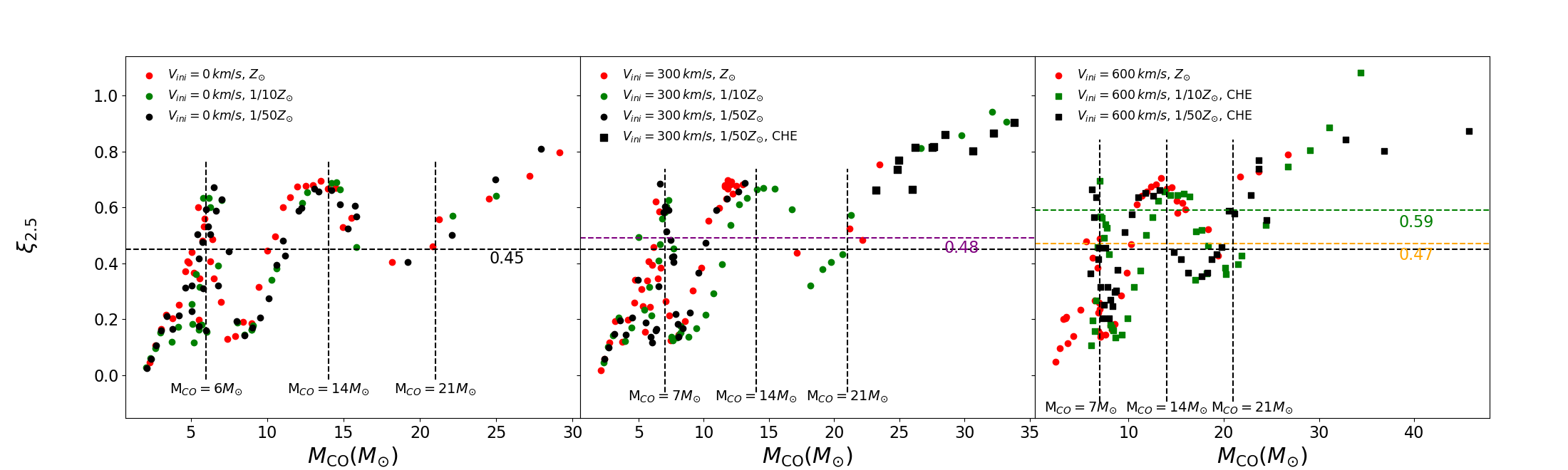}
    \end{minipage}
    \label{fig:fourimages}
    \caption{The $\xi_{\mathrm{2.5}}$ versus CO-core mass for models with varying initial rotational velocities $V_{\rm ini}$ and metallicities. The left, middle, and right panels correspond to models with $V_{\mathrm{ini}} = 0$, $300\,\mathrm{km\,s^{-1}}$, and $600\,\mathrm{km\,s^{-1}}$, respectively. Red, green, and black data points represent metallicities of $Z_{\odot}$, $1/10\,Z_{\odot}$, and $1/50\,Z_{\odot}$, respectively. Squares mark the occurrence of CHE.}
    \label{fig:6}
\end{figure*}

As discussed in the last subsection, the explodability of massive stars is determined by their metallicity, rotation, and specific ZAMS masses. In this section, we present the relationship between explodability criteria and ZAMS masses in our models.

Figure~\ref{fig:5} shows the pre-supernova star compactness parameter $\xi_{2.5}$ versus initial mass. Our results exhibit a similar trend to \cite{2023ApJ...952...79L} and \cite{2020ApJ...901..114A}. The $\xi_{2.5}$  of our current model does not exhibit a significant positive correlation with M$_{ZAMS}$, consistent with the simulated results reported in other studies \citep{1996ApJ...457..834T,2021A&A...645A...5S,2024A&A...682A.123T}. The final compactness, central entropy, iron-core mass, and binding energy, as functions of initial mass, follow similar patterns \citep{2014ApJ...783...10S,2023ApJ...945...19T,2021A&A...645A...5S,2024A&A...686A..45S,2024A&A...682A.123T}. \cite{2025A&A...695A..71L} investigated the underlying cause of this distinctive distribution. Taking the left panel of Fig.~\ref{fig:5} as an example, \cite{2025A&A...695A..71L} attributes the increasing compactness at \textasciitilde 10–20 M$_{\odot}$ and \textasciitilde 25–30 $M_{\odot}$ to the following reason: when neutrino energy loss exceeds the energy released by carbon/neon burning, the core contracts, driving the burning front outward. As the M$_{\rm ZAMS}$ increases, a larger carbon/neon-free core forms at the center, ultimately yielding a more massive and denser iron core with higher compactness. In contrast, stars of \textasciitilde 20–25 $M_{\odot}$ and \textasciitilde 30–35 $M_{\odot}$ experience declining compactness. \cite{2025A&A...695A..71L} argued that this drop is due to the earlier ignition of subsequent burning episodes, which slows down the progression of the carbon and oxygen burning fronts, caused by two factors: (1) the shortening of the main burning phase caused by fuel depletion and rising temperatures, and (2) the acceleration of core contraction when the fuel-depleted core exceeds the effective Chandrasekhar mass, leading to destabilization of electron degeneracy and accelerated collapse. These processes slow down core contraction, leading to reduced compactness. By examining the distribution of $\xi_{2.5}$ with respect to $M_{\rm ZAMS}$, we find that all model groups exhibit similar profiles. In the left and middle panels, the relation between M$_{\rm ZAMS}$ and $\xi_{\rm 2.5}$ is similar. In the right panel, the profile shifts toward lower mass regions as metallicity decreases. This shift arises because CHE occurs in the $1/10\,Z_{\odot}$ and $1/50\,Z_{\odot}$ model groups. Lower metallicity promotes more efficient CHE, which produces larger He cores and CO cores for the same M$_{\rm ZAMS}$ compared to higher metallicity models.

According to Figure~\ref{fig:5}, we use our explodability criterion to determine the initial mass range for FSN. Left panel ($V_{\mathrm{ini}} = 0$), adopting the explodability criterion of $\xi_{2.5} = 0.45$, stars with initial masses between approximately 21 and 23 $M_{\odot}$, between 30 and 41 $M_{\odot}$, and above approximately 50 $M_{\odot}$ can not undergo supernova explosions. These results are consistent with previous studies by \cite{2018ApJ...860...93S}, \cite{2012ApJ...757...69U}, and \cite{2021ApJ...911....6I}. For the middle panel ($V_{\mathrm{ini}} = 300~\mathrm{km\,s^{-1}}$), based on $\xi_{2.5}=0.48$, FSN forms in the initial mass ranges of approximately 20$\texttt{-}$22 $M_{\odot}$, 27$\texttt{-}$35 $M_{\odot}$, and above 40 $M_{\odot}$. For right panel ($V_{\rm ini}$ = 600 km s$^{-1}$), the results depend on metallicity: at solar metallicity ($ Z = Z_{\odot} $), the FSN initial mass range: \textasciitilde 14 $M_{\odot}$, 23$\texttt{-}$40 $M_{\odot}$ and above 50 $M_{\odot}$; for $1/10\,Z_{\odot}$, stars with $M_{\rm ZAMS}$ in the ranges of approximately \textasciitilde 11 $M_{\odot}$, 21--26 $M_{\odot}$ and above 40$M_{\odot}$ fail to undergo CCSN explosions; for $1/50\,Z_{\odot}$, stars with $M_{\rm ZAMS}$ in the ranges of approximately \textasciitilde 11 $M_{\odot}$, 17$\texttt{-}$20 $M_{\odot}$ and above 32 $M_{\odot}$ fail to undergo CCSN explosions. 

For models without CHE, the initial mass range for the occurrence of FSN differs slightly from the results obtained using the traditional criterion ($\xi_{2.5} = 0.45$). In contrast, pre-supernova star that have undergone CHE exhibit a reduced $M_{ZAMS}$ range for forming FSN when applying the $\xi_{2.5} = 0.59$ criterion. Due to their cores maintaining higher pre-supernova star angular velocities (see Fig~\ref{fig:3}), these systems - even when failing to produce successful CCSN explosions - are more likely to generate long gamma-ray bursts (lGRBs) \citep{1993ApJ...405..273W, 2006A&A...460..199Y}, and this results in luminous transients that may not conform to the observational definition of FSN \citep{2006A&A...460..199Y}. The production of lGRBs occurs when the specific angular ($j_{\rm co}$) momentum at any location within the CO core exceeds that of the last stable orbit ($j_{\rm Kerr,LSO}$) \citep{1972ApJ...178..347B}. CHE progenitors undergoing core collapse leave bare CO cores, and retain sufficient specific angular momentum during black hole formation to facilitate lGRB production \citep{2018ApJ...858..115A, 2020ApJ...901..114A,2023ApJ...952...79L}.

Figure~\ref{fig:6} shows the evolution of $\xi_{2.5}$ as a function of CO core mass at core collapse for different model groups. All model groups exhibit comparable trends: distinct compact regions of high $\xi_{2.5}$ appear at CO core masses of approximately 6–7 M$_{\odot}$, \textasciitilde 14 M$_{\odot}$, and $\gtrsim$ 21 M$_{\odot}$. These high-$\xi_{2.5}$ regions generally correspond to the domains favorable for failed supernova formation. Rotation and metallicity have minimal impact on the overall distribution of $\xi_{2.5}$ with CO core mass. However, models with CHE significantly narrow the CO core mass interval leading to failed supernovae (reducing the width of the peaks). It is noteworthy that the locations of the first two compact peaks observed in this study (at \textasciitilde 7 M$_{\odot}$ and \textasciitilde 14 M$_{\odot}$) are comparable to those reported by \cite{2024A&A...682A.123T}. That paper indicates that the location of the peak in the $\xi_{2.5}$ distribution versus CO core mass is affected by the convective core overshooting value.

\section{Conclusions}
Using the MESA stellar evolution code, we simulate stars with initial rotational velocities of $V_\mathrm{ini} = 0$, $300~\mathrm{km\,s^{-1}}$, and $600~\mathrm{km\,s^{-1}}$ at three metallicities ($Z_{\odot}$, $1/10\,Z_{\odot}$, and $1/50\,Z_{\odot}$), tracking their evolution from the ZAMS until iron core collapse at $1000~\mathrm{km\,s^{-1}}$. Using the MESA models of iron core collapse at 1000 km s$^{-1}$, we extract key parameters (enclosed mass, temperature, density, radial velocity, electron fraction, and angular velocity) as input parameters for GR1D, and subsequently simulate the CCSN phase with GR1D. We iteratively determine the critical value of \( f_{\mathrm{heat}} \) such that it lies within 1\% of the threshold required for a successful explosion. The corresponding time-averaged heating efficiency at this critical \( f_{\mathrm{heat}} \), denoted as \(\bar{\eta}_{\mathrm{heat}}^{\mathrm{crit}}\), is then used to evaluate whether the progenitor can explode in multidimensional simulations. Finally, we establish a correlation between the explosion outcomes and \(\xi_{2.5}\), and derive the critical compactness parameters $\xi_{2.5}$ for pre-supernova star explodability as follows: 0.45 for models with $V_{\text{ini}}$=0; 0.48 for the $V_{\text{ini}}$=300 km s$^{-1}$ group; 0.47 for $V_{\text{ini}}$=600 km s$^{-1}$ at $Z=Z_{\odot}$ ; and 0.59 for low metallicity (Z=1/10 and 1/50 $Z_\odot$). These criteria enable rapid assessment of pre-supernova stars explodability for EOS configurations resembling LS220. The upper limit of pre-supernova star compactness for producing CCSNe is significantly higher in models where massive stars have undergone CHE compared to those without CHE. This discrepancy primarily arises because the centrifugal force generated by rotational motion in pre-supernova star more effectively facilitates explosions compared to non-rotating scenarios. Traditional criteria ($\xi_{2.5}$=0.45) may be unsuitable when evaluating CHE progenitor explodability.

We established correlations between the explodability of stars in different model groups and their ZAMS masses, determining the initial ZAMS mass ranges for explode formation across these groups. For models without CHE, the ZAMS mass ranges leading to FSN exhibited little variation when compared with the ranges obtained using the $\xi_{2.5}<0.45$ explodability criterion. In contrast, model groups that underwent CHE showed a significantly increased initial ZAMS mass range for the occurrence of explosion, they are more likely to produce lGRBs, leading to a significantly reduced likelihood of FSN. Through analysis of the compactness parameter $\xi_{2.5}$ distribution as a function of CO core mass, we find that for model groups with otherwise identical initial parameters, rotation and metallicity show only limited influence on the distribution of compactness with CO-core mass.

\begin{acknowledgements}

We thank Evan O’Connor for helpful discussions and advice about using GR1D. This work received the support of the National Natural Science Foundation of China under grants 12373038, 12163005, U2031204 and 12288102; the Natural Science Foundation of Xinjiang No.2022TSYCLJ0006 and 2022D01D85.
\end{acknowledgements}
\bibliographystyle{aa}
\bibliography{aa55081-25}

\begin{thebibliography}{114}
\expandafter\ifx\csname natexlab\endcsname\relax\def\natexlab#1{#1}\fi

\bibitem[{{Aguilera-Dena} {et~al.}(2020){Aguilera-Dena}, {Langer},
  {Antoniadis}, \& {M{\"u}ller}}]{2020ApJ...901..114A}
{Aguilera-Dena}, D.~R., {Langer}, N., {Antoniadis}, J., \& {M{\"u}ller}, B.
  2020, \apj, 901, 114

\bibitem[{{Aguilera-Dena} {et~al.}(2022){Aguilera-Dena}, {Langer},
  {Antoniadis}, {Pauli}, {Dessart}, {Vigna-G{\'o}mez}, {Gr{\"a}fener}, \&
  {Yoon}}]{2022A&A...661A..60A}
{Aguilera-Dena}, D.~R., {Langer}, N., {Antoniadis}, J., {et~al.} 2022, \aap,
  661, A60

\bibitem[{{Aguilera-Dena} {et~al.}(2018){Aguilera-Dena}, {Langer}, {Moriya}, \&
  {Schootemeijer}}]{2018ApJ...858..115A}
{Aguilera-Dena}, D.~R., {Langer}, N., {Moriya}, T.~J., \& {Schootemeijer}, A.
  2018, \apj, 858, 115

\bibitem[{{Aguilera-Dena} {et~al.}(2023){Aguilera-Dena}, {M{\"u}ller},
  {Antoniadis}, {Langer}, {Dessart}, {Vigna-G{\'o}mez}, \&
  {Yoon}}]{2023A&A...671A.134A}
{Aguilera-Dena}, D.~R., {M{\"u}ller}, B., {Antoniadis}, J., {et~al.} 2023,
  \aap, 671, A134

\bibitem[{{Antoniadis} {et~al.}(2022){Antoniadis}, {Aguilera-Dena},
  {Vigna-G{\'o}mez}, {Kramer}, {Langer}, {M{\"u}ller}, {Tauris}, {Wang}, \&
  {Xu}}]{2022A&A...657L...6A}
{Antoniadis}, J., {Aguilera-Dena}, D.~R., {Vigna-G{\'o}mez}, A., {et~al.} 2022,
  \aap, 657, L6

\bibitem[{{Asplund} {et~al.}(2009){Asplund}, {Grevesse}, {Sauval}, \&
  {Scott}}]{2009ARA&A..47..481A}
{Asplund}, M., {Grevesse}, N., {Sauval}, A.~J., \& {Scott}, P. 2009, \araa, 47,
  481

\bibitem[{{Bardeen} {et~al.}(1972){Bardeen}, {Press}, \&
  {Teukolsky}}]{1972ApJ...178..347B}
{Bardeen}, J.~M., {Press}, W.~H., \& {Teukolsky}, S.~A. 1972, \apj, 178, 347

\bibitem[{{Bethe} \& {Wilson}(1985)}]{1985ApJ...295...14B}
{Bethe}, H.~A. \& {Wilson}, J.~R. 1985, \apj, 295, 14

\bibitem[{{Bjorkman} \& {Cassinelli}(1993)}]{1993ApJ...409..429B}
{Bjorkman}, J.~E. \& {Cassinelli}, J.~P. 1993, \apj, 409, 429

\bibitem[{{Boccioli} {et~al.}(2021){Boccioli}, {Mathews}, \&
  {O'Connor}}]{2021ApJ...912...29B}
{Boccioli}, L., {Mathews}, G.~J., \& {O'Connor}, E.~P. 2021, \apj, 912, 29

\bibitem[{{Boccioli} {et~al.}(2025){Boccioli}, {Vartanyan}, {O'Connor}, \&
  {Kasen}}]{2025MNRAS.tmp..931B}
{Boccioli}, L., {Vartanyan}, D., {O'Connor}, E.~P., \& {Kasen}, D. 2025, \mnras
  [\eprint[arXiv]{2501.06784}]

\bibitem[{{B{\"o}hm-Vitense}(1958)}]{1958ZA.....46..108B}
{B{\"o}hm-Vitense}, E. 1958, \zap, 46, 108

\bibitem[{{Brott} {et~al.}(2011){Brott}, {de Mink}, {Cantiello}, {Langer}, {de
  Koter}, {Evans}, {Hunter}, {Trundle}, \& {Vink}}]{2011A&A...530A.115B}
{Brott}, I., {de Mink}, S.~E., {Cantiello}, M., {et~al.} 2011, \aap, 530, A115

\bibitem[{{Bruenn} {et~al.}(2016){Bruenn}, {Lentz}, {Hix}, {Mezzacappa},
  {Harris}, {Messer}, {Endeve}, {Blondin}, {Chertkow}, {Lingerfelt},
  {Marronetti}, \& {Yakunin}}]{2016ApJ...818..123B}
{Bruenn}, S.~W., {Lentz}, E.~J., {Hix}, W.~R., {et~al.} 2016, \apj, 818, 123

\bibitem[{{Buras} {et~al.}(2006{\natexlab{a}}){Buras}, {Janka}, {Rampp}, \&
  {Kifonidis}}]{2006A&A...457..281B}
{Buras}, R., {Janka}, H.~T., {Rampp}, M., \& {Kifonidis}, K.
  2006{\natexlab{a}}, \aap, 457, 281

\bibitem[{{Buras} {et~al.}(2006{\natexlab{b}}){Buras}, {Rampp}, {Janka}, \&
  {Kifonidis}}]{2006A&A...447.1049B}
{Buras}, R., {Rampp}, M., {Janka}, H.~T., \& {Kifonidis}, K.
  2006{\natexlab{b}}, \aap, 447, 1049

\bibitem[{{Burrows} {et~al.}(2019){Burrows}, {Radice}, \&
  {Vartanyan}}]{2019MNRAS.485.3153B}
{Burrows}, A., {Radice}, D., \& {Vartanyan}, D. 2019, \mnras, 485, 3153

\bibitem[{{Burrows} {et~al.}(2020){Burrows}, {Radice}, {Vartanyan}, {Nagakura},
  {Skinner}, \& {Dolence}}]{2020MNRAS.491.2715B}
{Burrows}, A., {Radice}, D., {Vartanyan}, D., {et~al.} 2020, \mnras, 491, 2715

\bibitem[{{Burrows} {et~al.}(2006){Burrows}, {Reddy}, \&
  {Thompson}}]{2006NuPhA.777..356B}
{Burrows}, A., {Reddy}, S., \& {Thompson}, T.~A. 2006, \nphysa, 777, 356

\bibitem[{{Burrows} {et~al.}(2023){Burrows}, {Vartanyan}, \&
  {Wang}}]{2023ApJ...957...68B}
{Burrows}, A., {Vartanyan}, D., \& {Wang}, T. 2023, \apj, 957, 68

\bibitem[{{Burrows} {et~al.}(2024){Burrows}, {Wang}, \&
  {Vartanyan}}]{2024ApJ...964L..16B}
{Burrows}, A., {Wang}, T., \& {Vartanyan}, D. 2024, \apjl, 964, L16

\bibitem[{{Cardall} {et~al.}(2013){Cardall}, {Endeve}, \&
  {Mezzacappa}}]{2013PhRvD..87j3004C}
{Cardall}, C.~Y., {Endeve}, E., \& {Mezzacappa}, A. 2013, \prd, 87, 103004

\bibitem[{{Chan} {et~al.}(2018){Chan}, {M{\"u}ller}, {Heger}, {Pakmor}, \&
  {Springel}}]{2018ApJ...852L..19C}
{Chan}, C., {M{\"u}ller}, B., {Heger}, A., {Pakmor}, R., \& {Springel}, V.
  2018, \apjl, 852, L19

\bibitem[{{Colgate} \& {White}(1966)}]{1966ApJ...143..626C}
{Colgate}, S.~A. \& {White}, R.~H. 1966, \apj, 143, 626

\bibitem[{{Couch} {et~al.}(2020){Couch}, {Warren}, \&
  {O'Connor}}]{2020ApJ...890..127C}
{Couch}, S.~M., {Warren}, M.~L., \& {O'Connor}, E.~P. 2020, \apj, 890, 127

\bibitem[{{Diehl} {et~al.}(2021){Diehl}, {Lugaro}, {Heger}, {Sieverding},
  {Tang}, {Li}, {Li}, {Doherty}, {Krause}, {Wallner}, {Prantzos}, {Brinkman},
  {den Hartogh}, {Wehmeyer}, {Yag{\"u}e L{\'o}pez}, {Pleintinger}, {Banerjee},
  \& {Wang}}]{2021PASA...38...62D}
{Diehl}, R., {Lugaro}, M., {Heger}, A., {et~al.} 2021, \pasa, 38, e062

\bibitem[{{Ertl} {et~al.}(2016){Ertl}, {Janka}, {Woosley}, {Sukhbold}, \&
  {Ugliano}}]{2016ApJ...818..124E}
{Ertl}, T., {Janka}, H.~T., {Woosley}, S.~E., {Sukhbold}, T., \& {Ugliano}, M.
  2016, \apj, 818, 124

\bibitem[{{Farrell} {et~al.}(2020){Farrell}, {Groh}, {Meynet}, \&
  {Eldridge}}]{2020MNRAS.494L..53F}
{Farrell}, E.~J., {Groh}, J.~H., {Meynet}, G., \& {Eldridge}, J.~J. 2020,
  \mnras, 494, L53

\bibitem[{{Fischer} {et~al.}(2017){Fischer}, {Bastian}, {Blaschke}, {Cierniak},
  {Hempel}, {Kl{\"a}hn}, {Mart{\'\i}nez-Pinedo}, {Newton}, {R{\"o}pke}, \&
  {Typel}}]{2017PASA...34...67F}
{Fischer}, T., {Bastian}, N.-U., {Blaschke}, D., {et~al.} 2017, \pasa, 34, e067

\bibitem[{{Fischer} {et~al.}(2009){Fischer}, {Whitehouse}, {Mezzacappa},
  {Thielemann}, \& {Liebend{\"o}rfer}}]{2009A&A...499....1F}
{Fischer}, T., {Whitehouse}, S.~C., {Mezzacappa}, A., {Thielemann}, F.~K., \&
  {Liebend{\"o}rfer}, M. 2009, \aap, 499, 1

\bibitem[{{Fryer} \& {Warren}(2002)}]{2002ApJ...574L..65F}
{Fryer}, C.~L. \& {Warren}, M.~S. 2002, \apjl, 574, L65

\bibitem[{{Gogilashvili} {et~al.}(2023){Gogilashvili}, {Murphy}, \&
  {O'Connor}}]{2023MNRAS.524.4109G}
{Gogilashvili}, M., {Murphy}, J.~W., \& {O'Connor}, E.~P. 2023, \mnras, 524,
  4109

\bibitem[{{Halevi} {et~al.}(2023){Halevi}, {Wu}, {M{\"o}sta}, {Gottlieb},
  {Tchekhovskoy}, \& {Aguilera-Dena}}]{2023ApJ...944L..38H}
{Halevi}, G., {Wu}, B., {M{\"o}sta}, P., {et~al.} 2023, \apjl, 944, L38

\bibitem[{{Hamann} {et~al.}(1995){Hamann}, {Koesterke}, \&
  {Wessolowski}}]{1995A&A...299..151H}
{Hamann}, W.~R., {Koesterke}, L., \& {Wessolowski}, U. 1995, \aap, 299, 151

\bibitem[{{Heger} \& {Langer}(2000)}]{2000ApJ...544.1016H}
{Heger}, A. \& {Langer}, N. 2000, \apj, 544, 1016

\bibitem[{{Heger} {et~al.}(2000){Heger}, {Langer}, \&
  {Woosley}}]{2000ApJ...528..368H}
{Heger}, A., {Langer}, N., \& {Woosley}, S.~E. 2000, \apj, 528, 368

\bibitem[{{Heger} {et~al.}(2023){Heger}, {M{\"u}ller}, \&
  {Mandel}}]{2023arXiv230409350H}
{Heger}, A., {M{\"u}ller}, B., \& {Mandel}, I. 2023, arXiv e-prints,
  arXiv:2304.09350

\bibitem[{{Henneco} {et~al.}(2024){Henneco}, {Schneider}, \&
  {Laplace}}]{2024A&A...682A.169H}
{Henneco}, J., {Schneider}, F.~R.~N., \& {Laplace}, E. 2024, \aap, 682, A169

\bibitem[{{Herant} {et~al.}(1994){Herant}, {Benz}, {Hix}, {Fryer}, \&
  {Colgate}}]{1994ApJ...435..339H}
{Herant}, M., {Benz}, W., {Hix}, W.~R., {Fryer}, C.~L., \& {Colgate}, S.~A.
  1994, \apj, 435, 339

\bibitem[{{Horiuchi} {et~al.}(2011){Horiuchi}, {Beacom}, {Kochanek}, {Prieto},
  {Stanek}, \& {Thompson}}]{2011ApJ...738..154H}
{Horiuchi}, S., {Beacom}, J.~F., {Kochanek}, C.~S., {et~al.} 2011, \apj, 738,
  154

\bibitem[{{Horowitz}(2002)}]{2002PhRvD..65d3001H}
{Horowitz}, C.~J. 2002, \prd, 65, 043001

\bibitem[{{Ivanov} \& {Fern{\'a}ndez}(2021)}]{2021ApJ...911....6I}
{Ivanov}, M. \& {Fern{\'a}ndez}, R. 2021, \apj, 911, 6

\bibitem[{{Janka}(2001)}]{2001A&A...368..527J}
{Janka}, H.~T. 2001, \aap, 368, 527

\bibitem[{{Janka}(2012)}]{2012ARNPS..62..407J}
{Janka}, H.-T. 2012, Annual Review of Nuclear and Particle Science, 62, 407

\bibitem[{{Janka}(2017)}]{2017hsn..book.1095J}
{Janka}, H.-T. 2017, in Handbook of Supernovae, ed. A.~W. {Alsabti} \&
  P.~{Murdin}, 1095

\bibitem[{{Janka}(2025)}]{2025arXiv250214836J}
{Janka}, H.~T. 2025, arXiv e-prints, arXiv:2502.14836

\bibitem[{{Janka} {et~al.}(2016){Janka}, {Melson}, \&
  {Summa}}]{2016ARNPS..66..341J}
{Janka}, H.-T., {Melson}, T., \& {Summa}, A. 2016, Annual Review of Nuclear and
  Particle Science, 66, 341

\bibitem[{{Janka} \& {Mueller}(1996)}]{1996A&A...306..167J}
{Janka}, H.~T. \& {Mueller}, E. 1996, \aap, 306, 167

\bibitem[{{Kobulnicky} \& {Fryer}(2007)}]{2007ApJ...670..747K}
{Kobulnicky}, H.~A. \& {Fryer}, C.~L. 2007, \apj, 670, 747

\bibitem[{{Kuroda} {et~al.}(2018){Kuroda}, {Kotake}, {Takiwaki}, \&
  {Thielemann}}]{2018MNRAS.477L..80K}
{Kuroda}, T., {Kotake}, K., {Takiwaki}, T., \& {Thielemann}, F.-K. 2018,
  \mnras, 477, L80

\bibitem[{{Langer}(1992)}]{1992A&A...265L..17L}
{Langer}, N. 1992, \aap, 265, L17

\bibitem[{{Langer}(1998)}]{1998A&A...329..551L}
{Langer}, N. 1998, \aap, 329, 551

\bibitem[{{Laplace} {et~al.}(2021){Laplace}, {Justham}, {Renzo}, {G{\"o}tberg},
  {Farmer}, {Vartanyan}, \& {de Mink}}]{2021A&A...656A..58L}
{Laplace}, E., {Justham}, S., {Renzo}, M., {et~al.} 2021, \aap, 656, A58

\bibitem[{{Laplace} {et~al.}(2025){Laplace}, {Schneider}, \&
  {Podsiadlowski}}]{2025A&A...695A..71L}
{Laplace}, E., {Schneider}, F.~R.~N., \& {Podsiadlowski}, P. 2025, \aap, 695,
  A71

\bibitem[{{Lattimer} \& {Swesty}(1991)}]{1991NuPhA.535..331L}
{Lattimer}, J.~M. \& {Swesty}, D.~F. 1991, \nphysa, 535, 331

\bibitem[{{Lentz} {et~al.}(2015){Lentz}, {Bruenn}, {Hix}, {Mezzacappa},
  {Messer}, {Endeve}, {Blondin}, {Harris}, {Marronetti}, \&
  {Yakunin}}]{2015ApJ...807L..31L}
{Lentz}, E.~J., {Bruenn}, S.~W., {Hix}, W.~R., {et~al.} 2015, \apjl, 807, L31

\bibitem[{{Li} {et~al.}(2023){Li}, {Zhu}, {Guo}, {Liu}, \&
  {L{\"u}}}]{2023ApJ...952...79L}
{Li}, L., {Zhu}, C., {Guo}, S., {Liu}, H., \& {L{\"u}}, G. 2023, \apj, 952, 79

\bibitem[{{Li} {et~al.}(2011){Li}, {Leaman}, {Chornock}, {Filippenko},
  {Poznanski}, {Ganeshalingam}, {Wang}, {Modjaz}, {Jha}, {Foley}, \&
  {Smith}}]{2011MNRAS.412.1441L}
{Li}, W., {Leaman}, J., {Chornock}, R., {et~al.} 2011, \mnras, 412, 1441

\bibitem[{{Li} {et~al.}(2025){Li}, {Lu}, {L{\"u}}, {Zhu}, {Liu}, \&
  {Yu}}]{2025ApJ...979L..37L}
{Li}, Z., {Lu}, X., {L{\"u}}, G., {et~al.} 2025, \apjl, 979, L37

\bibitem[{{Liebend{\"o}rfer} {et~al.}(2005){Liebend{\"o}rfer}, {Rampp},
  {Janka}, \& {Mezzacappa}}]{2005ApJ...620..840L}
{Liebend{\"o}rfer}, M., {Rampp}, M., {Janka}, H.~T., \& {Mezzacappa}, A. 2005,
  \apj, 620, 840

\bibitem[{{Limongi} \& {Chieffi}(2018)}]{2018ApJS..237...13L}
{Limongi}, M. \& {Chieffi}, A. 2018, \apjs, 237, 13

\bibitem[{{Mabanta} \& {Murphy}(2018)}]{2018ApJ...856...22M}
{Mabanta}, Q.~A. \& {Murphy}, J.~W. 2018, \apj, 856, 22

\bibitem[{{Maeder} \& {Meynet}(1987)}]{1987A&A...182..243M}
{Maeder}, A. \& {Meynet}, G. 1987, \aap, 182, 243

\bibitem[{{Maltsev} {et~al.}(2025){Maltsev}, {Schneider}, {Mandel},
  {M{\"u}ller}, {Heger}, {R{\"o}pke}, \& {Laplace}}]{2025A&A...700A..20M}
{Maltsev}, K., {Schneider}, F.~R.~N., {Mandel}, I., {et~al.} 2025, \aap, 700,
  A20

\bibitem[{{Marchant} {et~al.}(2016){Marchant}, {Langer}, {Podsiadlowski},
  {Tauris}, \& {Moriya}}]{2016A&A...588A..50M}
{Marchant}, P., {Langer}, N., {Podsiadlowski}, P., {Tauris}, T.~M., \&
  {Moriya}, T.~J. 2016, \aap, 588, A50

\bibitem[{{Miller} {et~al.}(1993){Miller}, {Wilson}, \&
  {Mayle}}]{1993ApJ...415..278M}
{Miller}, D.~S., {Wilson}, J.~R., \& {Mayle}, R.~W. 1993, \apj, 415, 278

\bibitem[{{Moe} \& {Di Stefano}(2017)}]{2017ApJS..230...15M}
{Moe}, M. \& {Di Stefano}, R. 2017, \apjs, 230, 15

\bibitem[{{M{\"u}ller}(2015)}]{2015MNRAS.453..287M}
{M{\"u}ller}, B. 2015, \mnras, 453, 287

\bibitem[{{M{\"u}ller} {et~al.}(2016){M{\"u}ller}, {Heger}, {Liptai}, \&
  {Cameron}}]{2016MNRAS.460..742M}
{M{\"u}ller}, B., {Heger}, A., {Liptai}, D., \& {Cameron}, J.~B. 2016, \mnras,
  460, 742

\bibitem[{{M{\"u}ller} {et~al.}(2017){M{\"u}ller}, {Melson}, {Heger}, \&
  {Janka}}]{2017MNRAS.472..491M}
{M{\"u}ller}, B., {Melson}, T., {Heger}, A., \& {Janka}, H.-T. 2017, \mnras,
  472, 491

\bibitem[{{M{\"u}ller} {et~al.}(2019){M{\"u}ller}, {Tauris}, {Heger},
  {Banerjee}, {Qian}, {Powell}, {Chan}, {Gay}, \&
  {Langer}}]{2019MNRAS.484.3307M}
{M{\"u}ller}, B., {Tauris}, T.~M., {Heger}, A., {et~al.} 2019, \mnras, 484,
  3307

\bibitem[{{Murphy} \& {Burrows}(2008)}]{2008ApJ...688.1159M}
{Murphy}, J.~W. \& {Burrows}, A. 2008, \apj, 688, 1159

\bibitem[{{Obergaulinger} \& {Aloy}(2020)}]{2020MNRAS.492.4613O}
{Obergaulinger}, M. \& {Aloy}, M.~{\'A}. 2020, \mnras, 492, 4613

\bibitem[{{O'Connor}(2015)}]{2015ApJS..219...24O}
{O'Connor}, E. 2015, \apjs, 219, 24

\bibitem[{{O'Connor} {et~al.}(2018){O'Connor}, {Bollig}, {Burrows}, {Couch},
  {Fischer}, {Janka}, {Kotake}, {Lentz}, {Liebend{\"o}rfer}, {Messer},
  {Mezzacappa}, {Takiwaki}, \& {Vartanyan}}]{2018JPhG...45j4001O}
{O'Connor}, E., {Bollig}, R., {Burrows}, A., {et~al.} 2018, Journal of Physics
  G Nuclear Physics, 45, 104001

\bibitem[{{O'Connor} \& {Ott}(2010)}]{2010CQGra..27k4103O}
{O'Connor}, E. \& {Ott}, C.~D. 2010, Classical and Quantum Gravity, 27, 114103

\bibitem[{{O'Connor} \& {Ott}(2011)}]{2011ApJ...730...70O}
{O'Connor}, E. \& {Ott}, C.~D. 2011, \apj, 730, 70

\bibitem[{{Ott} {et~al.}(2008){Ott}, {Burrows}, {Dessart}, \&
  {Livne}}]{2008ApJ...685.1069O}
{Ott}, C.~D., {Burrows}, A., {Dessart}, L., \& {Livne}, E. 2008, \apj, 685,
  1069

\bibitem[{{Paxton} {et~al.}(2011){Paxton}, {Bildsten}, {Dotter}, {Herwig},
  {Lesaffre}, \& {Timmes}}]{2011ApJS..192....3P}
{Paxton}, B., {Bildsten}, L., {Dotter}, A., {et~al.} 2011, \apjs, 192, 3

\bibitem[{{Paxton} {et~al.}(2013){Paxton}, {Cantiello}, {Arras}, {Bildsten},
  {Brown}, {Dotter}, {Mankovich}, {Montgomery}, {Stello}, {Timmes}, \&
  {Townsend}}]{2013ApJS..208....4P}
{Paxton}, B., {Cantiello}, M., {Arras}, P., {et~al.} 2013, \apjs, 208, 4

\bibitem[{{Paxton} {et~al.}(2015){Paxton}, {Marchant}, {Schwab}, {Bauer},
  {Bildsten}, {Cantiello}, {Dessart}, {Farmer}, {Hu}, {Langer}, {Townsend},
  {Townsley}, \& {Timmes}}]{2015ApJS..220...15P}
{Paxton}, B., {Marchant}, P., {Schwab}, J., {et~al.} 2015, \apjs, 220, 15

\bibitem[{{Paxton} {et~al.}(2018){Paxton}, {Schwab}, {Bauer}, {Bildsten},
  {Blinnikov}, {Duffell}, {Farmer}, {Goldberg}, {Marchant}, {Sorokina},
  {Thoul}, {Townsend}, \& {Timmes}}]{2018ApJS..234...34P}
{Paxton}, B., {Schwab}, J., {Bauer}, E.~B., {et~al.} 2018, \apjs, 234, 34

\bibitem[{{Paxton} {et~al.}(2019){Paxton}, {Smolec}, {Schwab}, {Gautschy},
  {Bildsten}, {Cantiello}, {Dotter}, {Farmer}, {Goldberg}, {Jermyn}, {Kanbur},
  {Marchant}, {Thoul}, {Townsend}, {Wolf}, {Zhang}, \&
  {Timmes}}]{2019ApJS..243...10P}
{Paxton}, B., {Smolec}, R., {Schwab}, J., {et~al.} 2019, \apjs, 243, 10

\bibitem[{{Pejcha} \& {Thompson}(2015)}]{2015ApJ...801...90P}
{Pejcha}, O. \& {Thompson}, T.~A. 2015, \apj, 801, 90

\bibitem[{{Radice} {et~al.}(2017){Radice}, {Burrows}, {Vartanyan}, {Skinner},
  \& {Dolence}}]{2017ApJ...850...43R}
{Radice}, D., {Burrows}, A., {Vartanyan}, D., {Skinner}, M.~A., \& {Dolence},
  J.~C. 2017, \apj, 850, 43

\bibitem[{{Saio} {et~al.}(1988){Saio}, {Nomoto}, \&
  {Kato}}]{1988Natur.334..508S}
{Saio}, H., {Nomoto}, K., \& {Kato}, M. 1988, \nat, 334, 508

\bibitem[{{Sana} {et~al.}(2013){Sana}, {de Koter}, {de Mink}, {Dunstall},
  {Evans}, {H{\'e}nault-Brunet}, {Ma{\'\i}z Apell{\'a}niz},
  {Ram{\'\i}rez-Agudelo}, {Taylor}, {Walborn}, {Clark}, {Crowther}, {Herrero},
  {Gieles}, {Langer}, {Lennon}, \& {Vink}}]{2013A&A...550A.107S}
{Sana}, H., {de Koter}, A., {de Mink}, S.~E., {et~al.} 2013, \aap, 550, A107

\bibitem[{{Sana} {et~al.}(2012){Sana}, {de Mink}, {de Koter}, {Langer},
  {Evans}, {Gieles}, {Gosset}, {Izzard}, {Le Bouquin}, \&
  {Schneider}}]{2012Sci...337..444S}
{Sana}, H., {de Mink}, S.~E., {de Koter}, A., {et~al.} 2012, Science, 337, 444

\bibitem[{{Sana} {et~al.}(2014){Sana}, {Le Bouquin}, {Lacour}, {Berger},
  {Duvert}, {Gauchet}, {Norris}, {Olofsson}, {Pickel}, {Zins}, {Absil}, {de
  Koter}, {Kratter}, {Schnurr}, \& {Zinnecker}}]{2014ApJS..215...15S}
{Sana}, H., {Le Bouquin}, J.~B., {Lacour}, S., {et~al.} 2014, \apjs, 215, 15

\bibitem[{{Scheck} {et~al.}(2006){Scheck}, {Kifonidis}, {Janka}, \&
  {M{\"u}ller}}]{2006A&A...457..963S}
{Scheck}, L., {Kifonidis}, K., {Janka}, H.~T., \& {M{\"u}ller}, E. 2006, \aap,
  457, 963

\bibitem[{{Schneider} {et~al.}(2023){Schneider}, {Podsiadlowski}, \&
  {Laplace}}]{2023ApJ...950L...9S}
{Schneider}, F. R.~N., {Podsiadlowski}, P., \& {Laplace}, E. 2023, \apjl, 950,
  L9

\bibitem[{{Schneider} {et~al.}(2024){Schneider}, {Podsiadlowski}, \&
  {Laplace}}]{2024A&A...686A..45S}
{Schneider}, F.~R.~N., {Podsiadlowski}, P., \& {Laplace}, E. 2024, \aap, 686,
  A45

\bibitem[{{Schneider} {et~al.}(2021){Schneider}, {Podsiadlowski}, \&
  {M{\"u}ller}}]{2021A&A...645A...5S}
{Schneider}, F.~R.~N., {Podsiadlowski}, P., \& {M{\"u}ller}, B. 2021, \aap,
  645, A5

\bibitem[{{Shibata} {et~al.}(2011){Shibata}, {Kiuchi}, {Sekiguchi}, \&
  {Suwa}}]{2011PThPh.125.1255S}
{Shibata}, M., {Kiuchi}, K., {Sekiguchi}, Y., \& {Suwa}, Y. 2011, Progress of
  Theoretical Physics, 125, 1255

\bibitem[{{Smartt}(2015)}]{2015PASA...32...16S}
{Smartt}, S.~J. 2015, \pasa, 32, e016

\bibitem[{{Stegmann} {et~al.}(2022){Stegmann}, {Antonini}, \&
  {Moe}}]{2022MNRAS.516.1406S}
{Stegmann}, J., {Antonini}, F., \& {Moe}, M. 2022, \mnras, 516, 1406

\bibitem[{{Sukhbold} {et~al.}(2016){Sukhbold}, {Ertl}, {Woosley}, {Brown}, \&
  {Janka}}]{2016ApJ...821...38S}
{Sukhbold}, T., {Ertl}, T., {Woosley}, S.~E., {Brown}, J.~M., \& {Janka}, H.~T.
  2016, \apj, 821, 38

\bibitem[{{Sukhbold} \& {Woosley}(2014)}]{2014ApJ...783...10S}
{Sukhbold}, T. \& {Woosley}, S.~E. 2014, \apj, 783, 10

\bibitem[{{Sukhbold} {et~al.}(2018){Sukhbold}, {Woosley}, \&
  {Heger}}]{2018ApJ...860...93S}
{Sukhbold}, T., {Woosley}, S.~E., \& {Heger}, A. 2018, \apj, 860, 93

\bibitem[{{Summa} {et~al.}(2018){Summa}, {Janka}, {Melson}, \&
  {Marek}}]{2018ApJ...852...28S}
{Summa}, A., {Janka}, H.-T., {Melson}, T., \& {Marek}, A. 2018, \apj, 852, 28

\bibitem[{{Takahashi} {et~al.}(2023){Takahashi}, {Takiwaki}, \&
  {Yoshida}}]{2023ApJ...945...19T}
{Takahashi}, K., {Takiwaki}, T., \& {Yoshida}, T. 2023, \apj, 945, 19

\bibitem[{{Takiwaki} {et~al.}(2016){Takiwaki}, {Kotake}, \&
  {Suwa}}]{2016MNRAS.461L.112T}
{Takiwaki}, T., {Kotake}, K., \& {Suwa}, Y. 2016, \mnras, 461, L112

\bibitem[{{Temaj} {et~al.}(2024){Temaj}, {Schneider}, {Laplace}, {Wei}, \&
  {Podsiadlowski}}]{2024A&A...682A.123T}
{Temaj}, D., {Schneider}, F.~R.~N., {Laplace}, E., {Wei}, D., \&
  {Podsiadlowski}, P. 2024, \aap, 682, A123

\bibitem[{{Timmes} {et~al.}(1996){Timmes}, {Woosley}, \&
  {Weaver}}]{1996ApJ...457..834T}
{Timmes}, F.~X., {Woosley}, S.~E., \& {Weaver}, T.~A. 1996, \apj, 457, 834

\bibitem[{{Ugliano} {et~al.}(2012){Ugliano}, {Janka}, {Marek}, \&
  {Arcones}}]{2012ApJ...757...69U}
{Ugliano}, M., {Janka}, H.-T., {Marek}, A., \& {Arcones}, A. 2012, \apj, 757,
  69

\bibitem[{{Vartanyan} \& {Burrows}(2023)}]{2023MNRAS.526.5900V}
{Vartanyan}, D. \& {Burrows}, A. 2023, \mnras, 526, 5900

\bibitem[{{Vartanyan} {et~al.}(2021){Vartanyan}, {Laplace}, {Renzo},
  {G{\"o}tberg}, {Burrows}, \& {de Mink}}]{2021ApJ...916L...5V}
{Vartanyan}, D., {Laplace}, E., {Renzo}, M., {et~al.} 2021, \apjl, 916, L5

\bibitem[{{Vink} {et~al.}(2001){Vink}, {de Koter}, \&
  {Lamers}}]{2001A&A...369..574V}
{Vink}, J.~S., {de Koter}, A., \& {Lamers}, H.~J.~G.~L.~M. 2001, \aap, 369, 574

\bibitem[{{Wang} {et~al.}(2022){Wang}, {Vartanyan}, {Burrows}, \&
  {Coleman}}]{2022MNRAS.517..543W}
{Wang}, T., {Vartanyan}, D., {Burrows}, A., \& {Coleman}, M. S.~B. 2022,
  \mnras, 517, 543

\bibitem[{{Woosley}(1993)}]{1993ApJ...405..273W}
{Woosley}, S.~E. 1993, \apj, 405, 273

\bibitem[{{Woosley} {et~al.}(2020){Woosley}, {Sukhbold}, \&
  {Janka}}]{2020ApJ...896...56W}
{Woosley}, S.~E., {Sukhbold}, T., \& {Janka}, H.~T. 2020, \apj, 896, 56

\bibitem[{{Woosley} \& {Weaver}(1995)}]{1995ApJS..101..181W}
{Woosley}, S.~E. \& {Weaver}, T.~A. 1995, \apjs, 101, 181

\bibitem[{{Yoon} {et~al.}(2006){Yoon}, {Langer}, \&
  {Norman}}]{2006A&A...460..199Y}
{Yoon}, S.~C., {Langer}, N., \& {Norman}, C. 2006, \aap, 460, 199

\bibitem[{{Yoshida} {et~al.}(2021){Yoshida}, {Takiwaki}, {Aguilera-Dena},
  {Kotake}, {Takahashi}, {Nakamura}, {Umeda}, \&
  {Langer}}]{2021MNRAS.506L..20Y}
{Yoshida}, T., {Takiwaki}, T., {Aguilera-Dena}, D.~R., {et~al.} 2021, \mnras,
  506, L20

\end{thebibliography}
\begin{appendix}
\section{Grid simulations at higher resolution}
We set a high-resolution grid comprising 1200 radial zones. Within the inner 20 km, 200 radial zones are configured at a fine spacing of 100 m. Beyond 20 km, the grid extends logarithmically to an outer boundary where the density reaches 2000 g cm$^{-3}$.

\begin{figure*}[htbp]
    \centering

    \begin{minipage}[b]{0.35\textwidth}
        \centering
        \includegraphics[width=\textwidth]{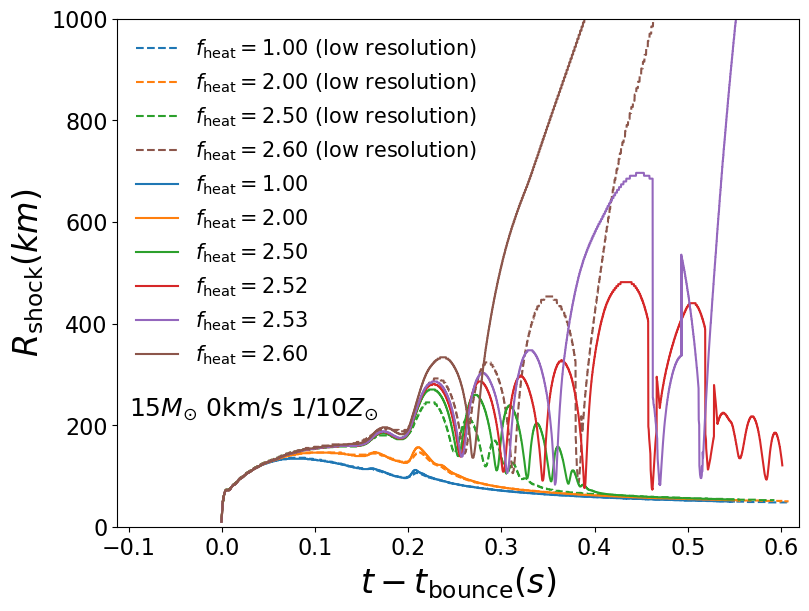}
    \end{minipage}
    \hspace{0.01\textwidth}
    \begin{minipage}[b]{0.35\textwidth}
        \centering
        \includegraphics[width=\textwidth]{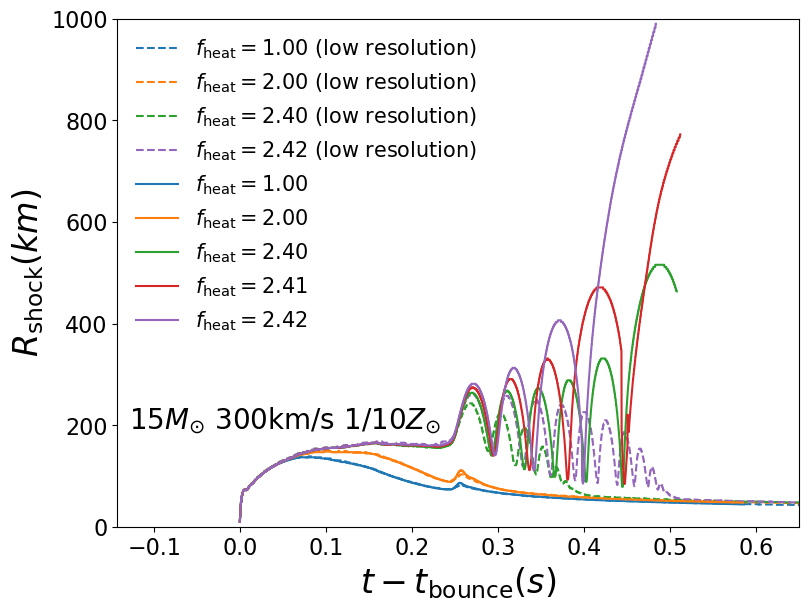}
    \end{minipage}

    \vspace{0.5cm}

    \begin{minipage}[b]{0.35\textwidth}
        \centering
        \includegraphics[width=\textwidth]{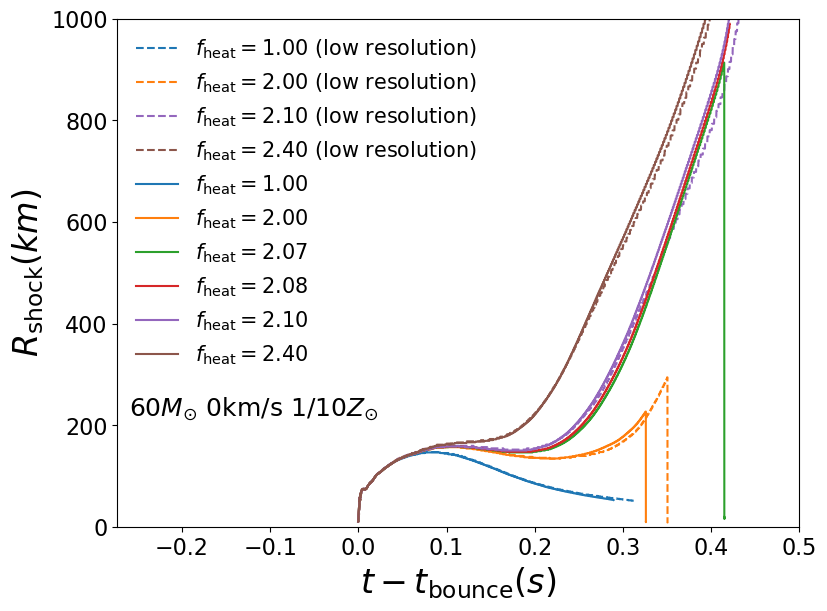}
    \end{minipage}
    \hspace{0.01\textwidth}
    \begin{minipage}[b]{0.35\textwidth}
        \centering
        \includegraphics[width=\textwidth]{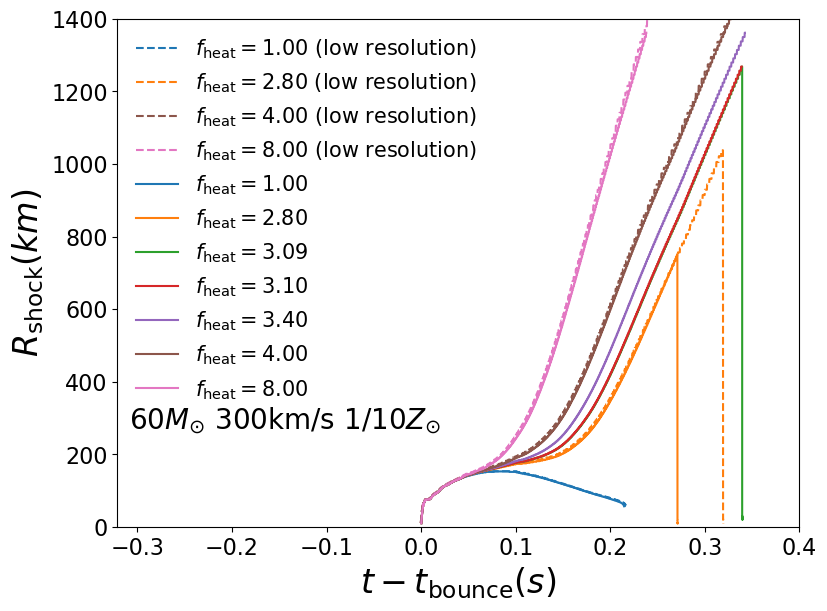}
    \end{minipage}

    \caption{The shock wave radius evolution versus explosion time $t_{\mathrm{bounce}}$ for different $f_{\mathrm{heat}}$ values. Simulations start with an iron core collapse velocity of 1000 km s$^{-1}$, where $t=0$ corresponds to shock wave formation. Dashed curves depict simulations employing coarse numerical grids, while solid lines correspond to higher-resolution implementations. Top and bottom panels show 15 M$_{\odot}$ and 60 M$_{\odot}$ ZAMS masses respectively, with initial velocities $V_{\mathrm{ini}} = 0$ (left panels) and 300 km s$^{-1}$ (right panels). All models have $1/10$ Z$_{\odot}$ metallicity. }
    \label{fig:7}
\end{figure*}

Figure~\ref{fig:7} shows the shock radius evolution with time $t_{\mathrm{bounce}}$ for different $f_{\mathrm{heat}}$ values. (The four pre-supernova star models simulated in here are identical to those presented in Figures 2 and 3 of the main text). Comparisons of simulations at different grid resolutions indicate that at low $f_{\rm heat}$ values, results from both grid resolutions are nearly identical. As $f_{\text{heat}}$ increases, simulations remain consistent until the early shock stagnation phase; beyond this point, however, the finer grid demonstrates a higher likelihood of shock revival and explosion. Simulations of the identical initial setup performed at varying grid resolutions yield deviations in $f_{\text{heat}}^{\text{crit}}$ that average less than $ 3\%$.

\begin{figure*}[htbp]
    \centering

    \begin{minipage}[b]{0.35\textwidth}
        \centering
        \includegraphics[width=\textwidth]{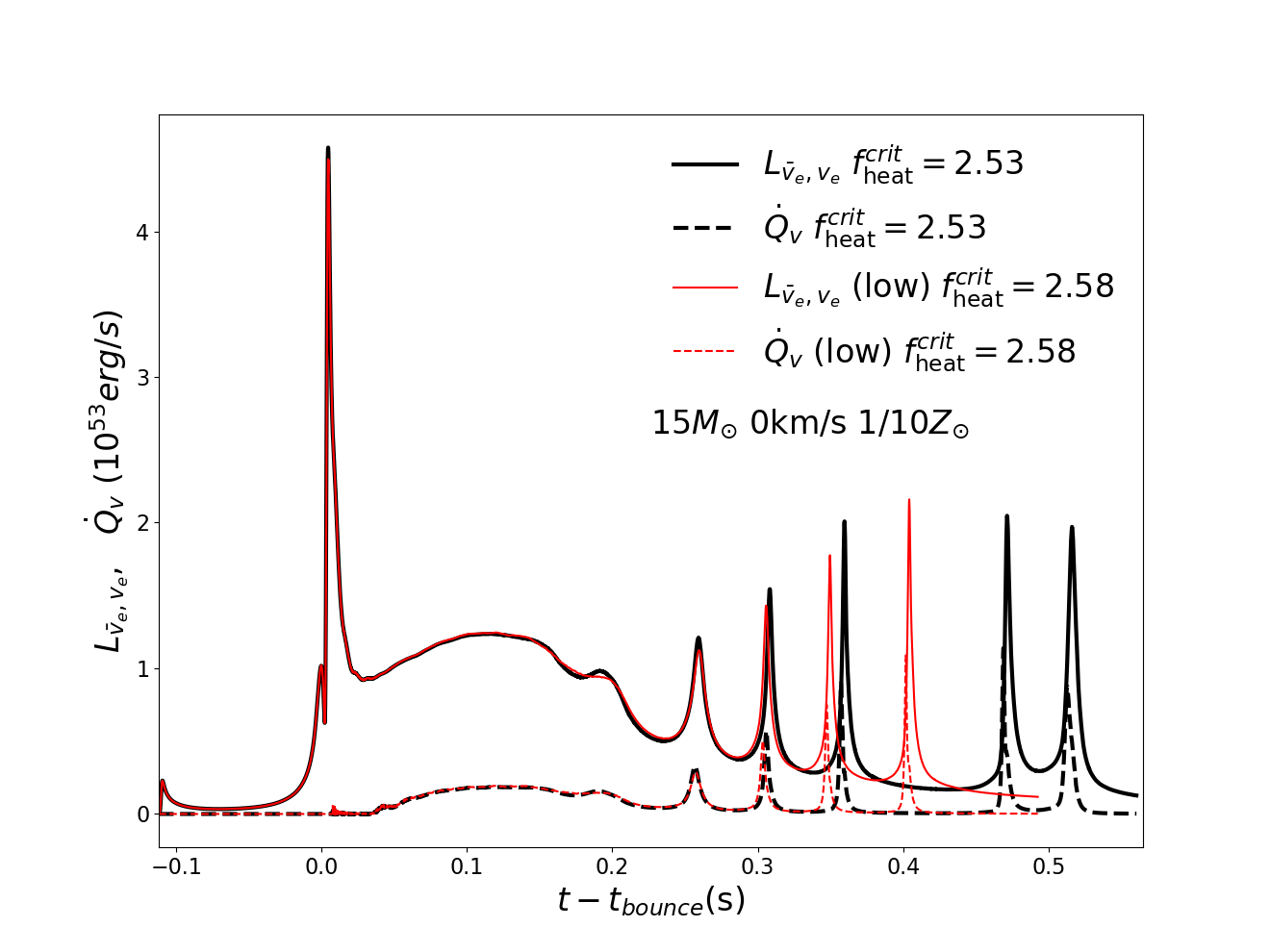}
    \end{minipage}
    \hspace{0.01\textwidth}
    \begin{minipage}[b]{0.35\textwidth}
        \centering
        \includegraphics[width=\textwidth]{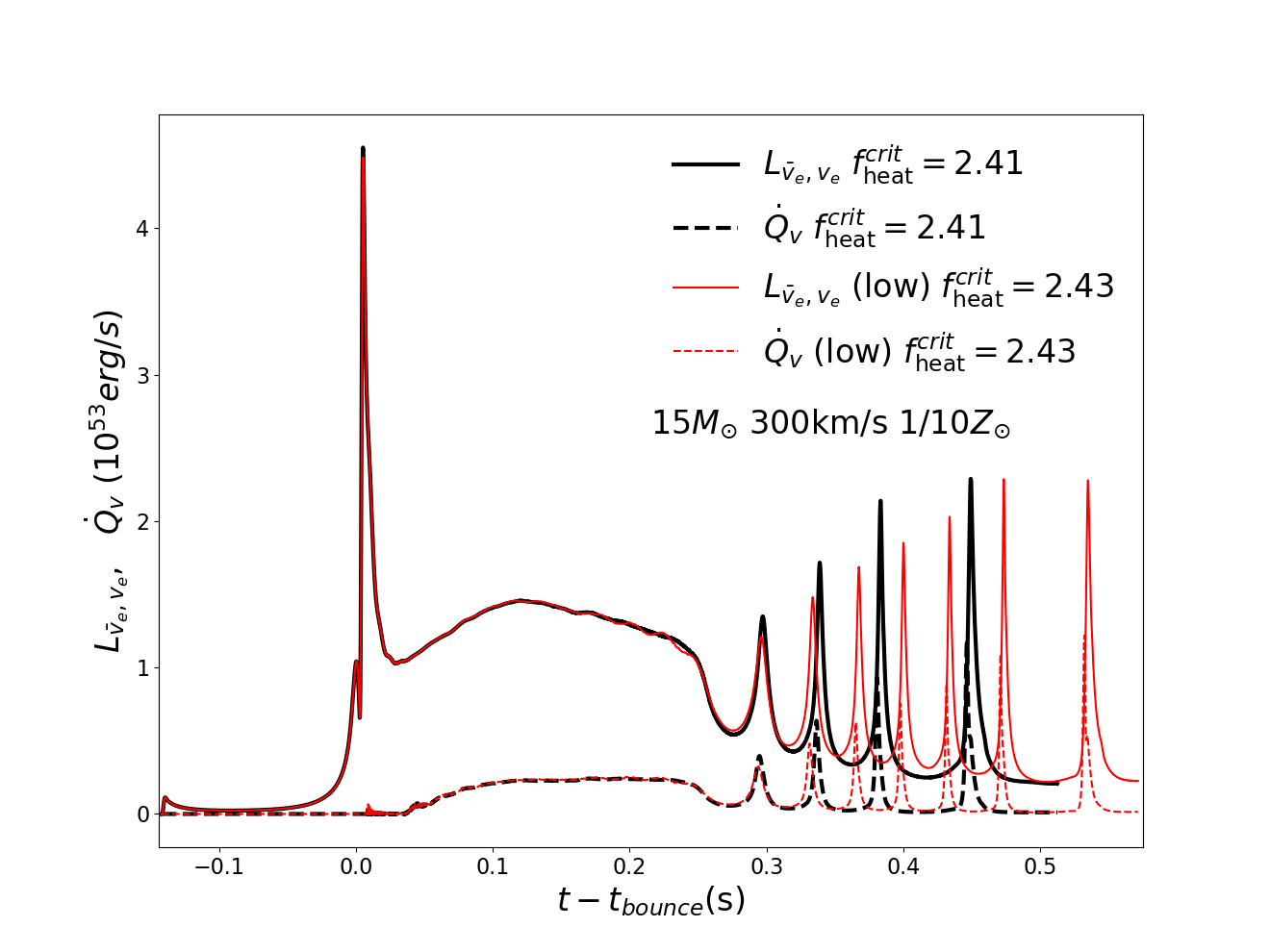}
    \end{minipage}

    \vspace{0.5cm}

    \begin{minipage}[b]{0.35\textwidth}
        \centering
        \includegraphics[width=\textwidth]{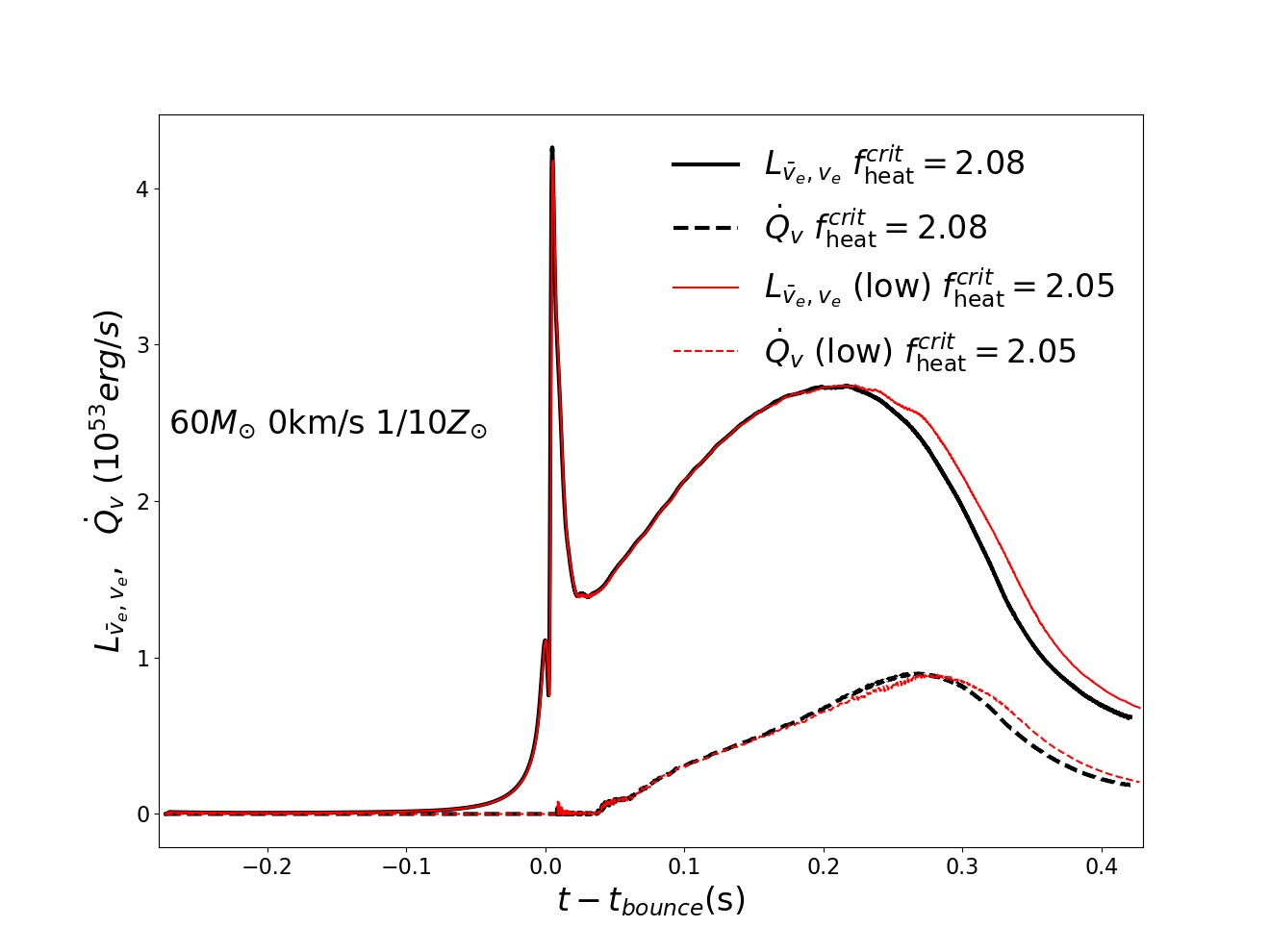}
    \end{minipage}
    \hspace{0.01\textwidth}
    \begin{minipage}[b]{0.35\textwidth}
        \centering
        \includegraphics[width=\textwidth]{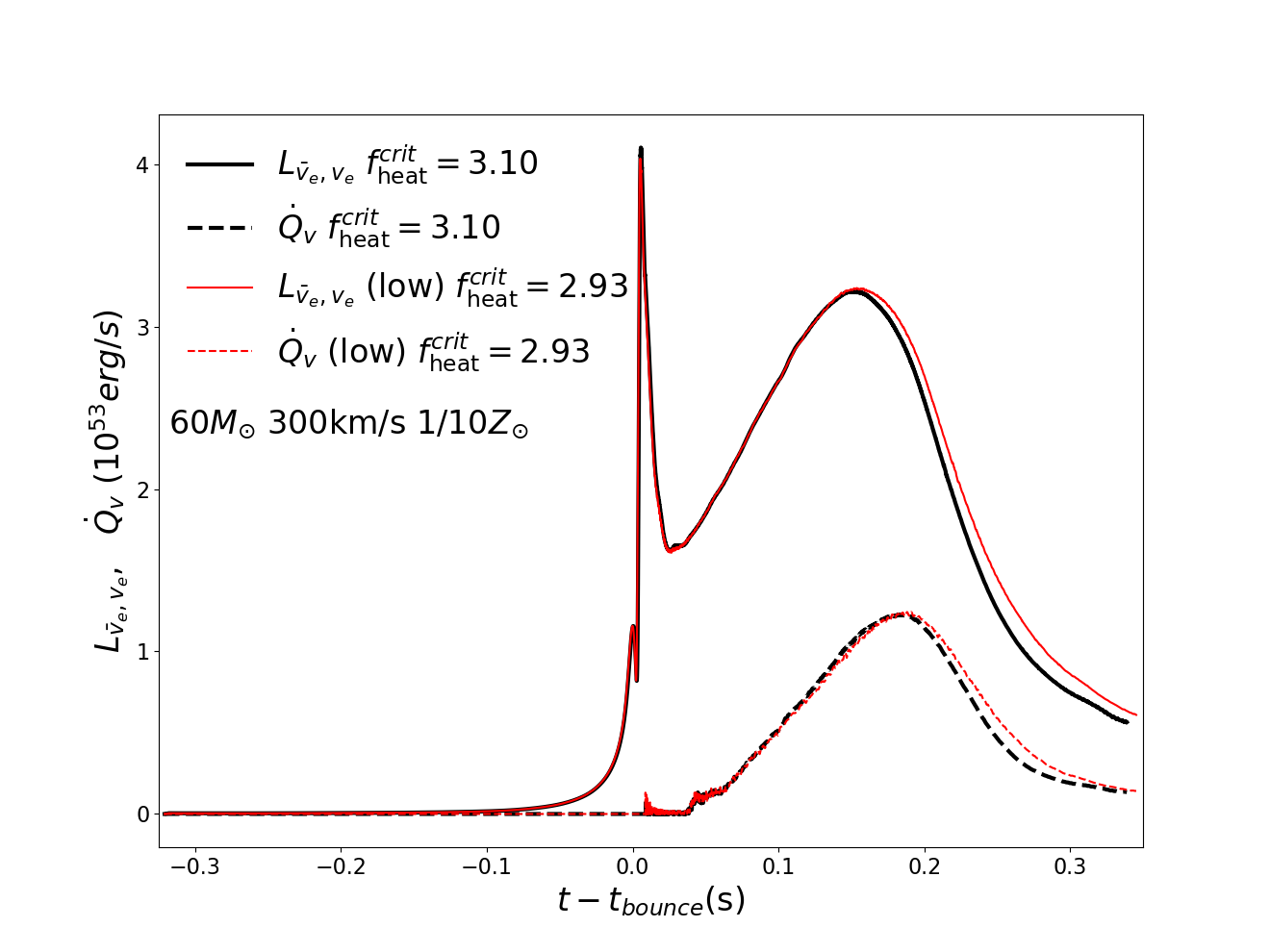}
    \end{minipage}
    \caption{For $f_{\mathrm{heat}} = f_{\mathrm{heat}}^{\mathrm{crit}}$, evolution of electron neutrino and antineutrino luminosities $L_{\bar{\nu}_e,\nu_e}$ (solid lines) at the gain radius and net neutrino energy deposition rate in the gain layer $\dot{Q}_{\nu}$ ($\int_{\text{gain}} \dot{q}^{+}_{v} \, dV$, dashed lines) versus time $t_{\mathrm{bounce}}$. Simulations start with an iron core collapse velocity of 1000 km s$^{-1}$, where $t=0$ corresponds to shock wave formation. Red lines depict simulations employing coarse numerical grids, while black curves correspond to higher-resolution implementations. Top and bottom panels show 15 M$_{\odot}$ and 60 M$_{\odot}$ ZAMS masses respectively, with initial velocities $V_{\mathrm{ini}} = 0$ (left panels) and 300 km s$^{-1}$ (right panels). All models have $1/10$ Z$_{\odot}$ metallicity.}
    \label{fig:8}
    
\end{figure*}

Figure~\ref{fig:8} depicts the temporal evolution of $L_{\bar{\nu}_e,\nu_e}$ at the gain radius and $\dot{Q}_{\nu}$ in the critical model across different grid resolutions. We calculate the critical heating efficiency parameters $\bar{\eta}_{\mathrm{heat}}^{\mathrm{crit}}$ for the 1200 grid configuration, with the upper panel (left to right) corresponding to 0.109 and 0.129, and the lower panel (left to right) to 0.246 and 0.253.

The results indicate a mean relative deviation of $\sim 3\%$ in $\bar{\eta}_{\mathrm{heat}}^{\mathrm{crit}}$ between grid configurations. Given that simulations with high-resolution grids require over five times the computational resources of their low-resolution counterparts, we employed the low-resolution setup (600 total cells) for all models.

\end{appendix}
\end{document}